\documentclass[journal]{IEEEtran}
\usepackage{cite}
\usepackage{amsmath,amssymb,amsfonts,bm}
\usepackage[linesnumbered,ruled,vlined]{algorithm2e}
\usepackage{mathrsfs}
\usepackage{amsthm}
\usepackage{dsfont}
\usepackage{algorithmic}
\usepackage{subfigure}
\usepackage{graphicx}
\usepackage{textcomp}
 \usepackage{threeparttable}
\usepackage{xcolor}
\usepackage{bbm}
\usepackage[nolist,withpage]{acronym}
\usepackage{array}
\usepackage{tikz}
\usetikzlibrary{spy}
\usepackage{pgfplots}
\pgfplotsset{compat=newest}

\definecolor{copperrose}{rgb}{0.6, 0.4, 0.4}
\definecolor{azure}{rgb}{0.0, 0.5, 1.0}
\definecolor{ashgrey}{rgb}{0.7, 0.75, 0.71}
\definecolor{chestnut}{rgb}{0.8, 0.36, 0.36}
\definecolor{airforceblue}{rgb}{0.36, 0.54, 0.66}
\definecolor{cadmiumorange}{rgb}{0.93, 0.53, 0.18}
\definecolor{bleudefrance}{rgb}{0.19, 0.55, 0.91}
\definecolor{carolinablue}{rgb}{0.6, 0.73, 0.89}
\definecolor{blue(ncs)}{rgb}{0.0, 0.53, 0.74}
\definecolor{dodgerblue}{rgb}{0.12, 0.56, 1.0}
\definecolor{cssgreen}{rgb}{0.0, 0.5, 0.0}
\definecolor{cadmiumgreen}{rgb}{0.0, 0.42, 0.24}
\definecolor{cadmiumorange}{rgb}{0.93, 0.53, 0.18}
\definecolor{amaranth}{rgb}{0.9, 0.17, 0.31}
\definecolor{bluegray}{rgb}{0.4, 0.6, 0.8}
\definecolor{cerulean}{rgb}{0.0, 0.48, 0.65}
\definecolor{ceil}{rgb}{0.57, 0.63, 0.81}

\usepackage{etoolbox}
\makeatletter
\newif\if@in@acrolist
\AtBeginEnvironment{acronym}{\@in@acrolisttrue}
\newrobustcmd{\LU}[2]{\if@in@acrolist#1\else#2\fi}

\newcommand{\ACF}[1]{{\@in@acrolisttrue\acf{#1}}}

\begin{document}


\begin{acronym}[LTE-Advanced]
  \acro{2G}{Second Generation}
  \acro{3-DAP}{3-Dimensional Assignment Problem}
  \acro{3G}{3$^\text{rd}$~Generation}
  \acro{3GPP}{3$^\text{rd}$~Generation Partnership Project}
  \acro{4G}{4$^\text{th}$~Generation}
  \acro{5G}{5$^\text{th}$~Generation}
  \acro{AA}{Antenna Array}
  \acro{AC}{Admission Control}
  \acro{AD}{Attack-Decay}
  \acro{ADC}{analog-to-digital converter}
  \acro{ADMM}{alternating direction method of multipliers}
  \acro{ADSL}{Asymmetric Digital Subscriber Line}
  \acro{AHW}{Alternate Hop-and-Wait}
  \acro{AirComp}{over-the-air computation}
  \acro{AMC}{Adaptive Modulation and Coding}
  \acro{ANN}{artificial neural network}
  \acro{AP}{\LU{A}{a}ccess \LU{P}{p}oint}
  \acro{APA}{Adaptive Power Allocation}
  \acro{ARMA}{Autoregressive Moving Average}
  \acro{ARQ}{\LU{A}{a}utomatic \LU{R}{r}epeat \LU{R}{r}equest}
  \acro{ASK}{Amplitude Shift Keying}
  \acro{ATES}{Adaptive Throughput-based Efficiency-Satisfaction Trade-Off}
  \acro{AWGN}{additive white Gaussian noise}
  \acro{BAA}{\LU{B}{b}roadband \LU{A}{a}nalog \LU{A}{a}ggregation}
  \acro{BB}{Branch and Bound}
  \acro{BCD}{block coordinate descent}
  \acro{BD}{Block Diagonalization}
  \acro{BER}{Bit Error Rate}
  \acro{BF}{Best Fit}
  \acro{BFD}{bidirectional full duplex}
  \acro{BLER}{BLock Error Rate}
  \acro{BPC}{Binary Power Control}
  \acro{BPSK}{Binary Phase-Shift Keying}
  \acro{BRA}{Balanced Random Allocation}
  \acro{BS}{base station}
  \acro{BSUM}{block successive upper-bound minimization}
  \acro{CAP}{Combinatorial Allocation Problem}
  \acro{CAPEX}{Capital Expenditure}
  \acro{CBF}{Coordinated Beamforming}
  \acro{CBR}{Constant Bit Rate}
  \acro{CBS}{Class Based Scheduling}
  \acro{CC}{Congestion Control}
  \acro{CDF}{Cumulative Distribution Function}
  \acro{CDMA}{Code-Division Multiple Access}
  \acro{CE}{\LU{C}{c}hannel \LU{E}{e}stimation}
  \acro{CFO}{carrier frequency offset}
  \acro{CL}{Closed Loop}
  \acro{CLPC}{Closed Loop Power Control}
  \acro{CML}{centralized machine learning}
  \acro{CNR}{Channel-to-Noise Ratio}
  \acro{CNN}{\LU{C}{c}onvolutional \LU{N}{n}eural \LU{N}{n}etwork}
  \acro{CP}{computation point}
  \acro{CPA}{Cellular Protection Algorithm}
  \acro{CPICH}{Common Pilot Channel}
  \acro{CoCoA}{\LU{C}{c}ommunication efficient distributed dual \LU{C}{c}oordinate \LU{A}{a}scent}
  \acro{CoMAC}{\LU{C}{c}omputation over \LU{M}{m}ultiple-\LU{A}{a}ccess \LU{C}{c}hannels}
  \acro{CoMP}{Coordinated Multi-Point}
  \acro{CQI}{Channel Quality Indicator}
  \acro{CRM}{Constrained Rate Maximization}
	\acro{CRN}{Cognitive Radio Network}
  \acro{CS}{Coordinated Scheduling}
  \acro{CSI}{\LU{C}{c}hannel \LU{S}{s}tate \LU{I}{i}nformation}
  \acro{CSMA}{\LU{C}{c}arrier \LU{S}{s}ense \LU{M}{m}ultiple \LU{A}{a}ccess}
  \acro{CUE}{Cellular User Equipment}
  \acro{D2D}{device-to-device}
  \acro{DAC}{digital-to-analog converter}
  \acro{DC}{direct current}
  \acro{DCA}{Dynamic Channel Allocation}
  \acro{DE}{Differential Evolution}
  \acro{DFT}{Discrete Fourier Transform}
  \acro{DIST}{Distance}
  \acro{DL}{downlink}
  \acro{DMA}{Double Moving Average}
  \acro{DML}{Distributed Machine Learning}
  \acro{DMRS}{demodulation reference signal}
  \acro{D2DM}{D2D Mode}
  \acro{DMS}{D2D Mode Selection}
  \acro{DPC}{Dirty Paper Coding}
  \acro{DRA}{Dynamic Resource Assignment}
  \acro{DSA}{Dynamic Spectrum Access}
  \acro{DSGD}{\LU{D}{d}istributed \LU{S}{s}tochastic \LU{G}{g}radient \LU{D}{d}escent}
  \acro{DSM}{Delay-based Satisfaction Maximization}
  \acro{ECC}{Electronic Communications Committee}
  \acro{EFLC}{Error Feedback Based Load Control}
  \acro{EI}{Efficiency Indicator}
  \acro{eNB}{Evolved Node B}
  \acro{EPA}{Equal Power Allocation}
  \acro{EPC}{Evolved Packet Core}
  \acro{EPS}{Evolved Packet System}
  \acro{E-UTRAN}{Evolved Universal Terrestrial Radio Access Network}
  \acro{ES}{Exhaustive Search}
  \acro{FC}{\LU{F}{f}usion \LU{C}{c}enter}
  \acro{FD}{\LU{F}{f}ederated \LU{D}{d}istillation}
  \acro{FDD}{frequency division duplex}
  \acro{FDM}{Frequency Division Multiplexing}
  \acro{FDMA}{\LU{F}{f}requency \LU{D}{d}ivision \LU{M}{m}ultiple \LU{A}{a}ccess}
  \acro{FedAvg}{\LU{F}{f}ederated \LU{A}{a}veraging}
  \acro{FER}{Frame Erasure Rate}
  \acro{FF}{Fast Fading}
  \acro{FL}{federated learning}
  \acro{FSB}{Fixed Switched Beamforming}
  \acro{FST}{Fixed SNR Target}
  \acro{FTP}{File Transfer Protocol}
  \acro{GA}{Genetic Algorithm}
  \acro{GBR}{Guaranteed Bit Rate}
  \acro{GLR}{Gain to Leakage Ratio}
  \acro{GOS}{Generated Orthogonal Sequence}
  \acro{GPL}{GNU General Public License}
  \acro{GRP}{Grouping}
  \acro{HARQ}{Hybrid Automatic Repeat Request}
  \acro{HD}{half-duplex}
  \acro{HMS}{Harmonic Mode Selection}
  \acro{HOL}{Head Of Line}
  \acro{HSDPA}{High-Speed Downlink Packet Access}
  \acro{HSPA}{High Speed Packet Access}
  \acro{HTTP}{HyperText Transfer Protocol}
  \acro{ICMP}{Internet Control Message Protocol}
  \acro{ICI}{Intercell Interference}
  \acro{ID}{Identification}
  \acro{IETF}{Internet Engineering Task Force}
  \acro{ILP}{Integer Linear Program}
  \acro{JRAPAP}{Joint RB Assignment and Power Allocation Problem}
  \acro{UID}{Unique Identification}
  \acro{IID}{\LU{I}{i}ndependent and \LU{I}{i}dentically \LU{D}{d}istributed}
  \acro{IIR}{Infinite Impulse Response}
  \acro{ILP}{Integer Linear Problem}
  \acro{IMT}{International Mobile Telecommunications}
  \acro{INV}{Inverted Norm-based Grouping}
  \acro{IoT}{Internet of Things}
  \acro{IP}{Integer Programming}
  \acro{IPv6}{Internet Protocol Version 6}
  \acro{ISD}{Inter-Site Distance}
  \acro{ISI}{Inter Symbol Interference}
  \acro{ITU}{International Telecommunication Union}
  \acro{JAFM}{joint assignment and fairness maximization}
  \acro{JAFMA}{joint assignment and fairness maximization algorithm}
  \acro{JOAS}{Joint Opportunistic Assignment and Scheduling}
  \acro{JOS}{Joint Opportunistic Scheduling}
  \acro{JP}{Joint Processing}
	\acro{JS}{Jump-Stay}
  \acro{KKT}{Karush-Kuhn-Tucker}
  \acro{L3}{Layer-3}
  \acro{LAC}{Link Admission Control}
  \acro{LA}{Link Adaptation}
  \acro{LC}{Load Control}
  \acro{LDC}{\LU{L}{l}earning-\LU{D}{d}riven \LU{C}{c}ommunication}
  \acro{LDPC}{low-density parity-check}
  \acro{LOS}{line of sight}
  \acro{LP}{linear programming}
  \acro{LTE}{Long Term Evolution}
	\acro{LTE-A}{\ac{LTE}-Advanced}
  \acro{LTE-Advanced}{Long Term Evolution Advanced}
  \acro{M2M}{Machine-to-Machine}
  \acro{MAC}{multiple access computing}
  \acro{MANET}{Mobile Ad hoc Network}
  \acro{MC}{Modular Clock}
  \acro{MCS}{Modulation and Coding Scheme}
  \acro{MDB}{Measured Delay Based}
  \acro{MDI}{Minimum D2D Interference}
  \acro{MF}{Matched Filter}
  \acro{MG}{Maximum Gain}
  \acro{MH}{Multi-Hop}
  \acro{MIMO}{\LU{M}{m}ultiple \LU{I}{i}nput \LU{M}{m}ultiple \LU{O}{o}utput}
  \acro{MINLP}{mixed integer nonlinear programming}
  \acro{MIP}{mixed integer programming}
  \acro{MIQCP}{mixed integer quadratic constrained programming}
  \acro{MISO}{multiple input single output}
  \acro{ML}{machine learning}
  \acro{MLE}{maximum likelihood estimator}
  \acro{MLWDF}{Modified Largest Weighted Delay First}
  \acro{MME}{Mobility Management Entity}
  \acro{MMSE}{minimum mean squared error}
  \acro{MOS}{Mean Opinion Score}
  \acro{MPF}{Multicarrier Proportional Fair}
  \acro{MRA}{Maximum Rate Allocation}
  \acro{MR}{Maximum Rate}
  \acro{MRC}{Maximum Ratio Combining}
  \acro{MRT}{Maximum Ratio Transmission}
  \acro{MRUS}{Maximum Rate with User Satisfaction}
  \acro{MS}{Mode Selection}
  \acro{MSE}{\LU{M}{m}ean \LU{S}{s}quared \LU{E}{e}rror}
  \acro{MSI}{Multi-Stream Interference}
  \acro{MTC}{Machine-Type Communication}
  \acro{MTSI}{Multimedia Telephony Services over IMS}
  \acro{MTSM}{Modified Throughput-based Satisfaction Maximization}
  \acro{MU-MIMO}{Multi-User Multiple Input Multiple Output}
  \acro{MU}{Multi-User}
  \acro{NAS}{Non-Access Stratum}
  \acro{NB}{Node B}
	\acro{NCL}{Neighbor Cell List}
  \acro{NLP}{Nonlinear Programming}
  \acro{NLOS}{non-line of sight}
  \acro{NMSE}{normalized mean square error}
  \acro{NOMA}{\LU{N}{n}on-\LU{O}{o}rthogonal \LU{M}{m}ultiple \LU{A}{a}ccess}
  \acro{NORM}{Normalized Projection-based Grouping}
  \acro{NP}{non-polynomial time}
  \acro{NRT}{Non-Real Time}
  \acro{NSPS}{National Security and Public Safety Services}
  \acro{O2I}{Outdoor to Indoor}
  \acro{OFDMA}{\LU{O}{o}rthogonal \LU{F}{f}requency \LU{D}{d}ivision \LU{M}{m}ultiple \LU{A}{a}ccess}
  \acro{OFDM}{Orthogonal Frequency Division Multiplexing}
  \acro{OFPC}{Open Loop with Fractional Path Loss Compensation}
	\acro{O2I}{Outdoor-to-Indoor}
  \acro{OL}{Open Loop}
  \acro{OLPC}{Open-Loop Power Control}
  \acro{OL-PC}{Open-Loop Power Control}
  \acro{OPEX}{Operational Expenditure}
  \acro{ORB}{Orthogonal Random Beamforming}
  \acro{JO-PF}{Joint Opportunistic Proportional Fair}
  \acro{OSI}{Open Systems Interconnection}
  \acro{PAIR}{D2D Pair Gain-based Grouping}
  \acro{PAM}{pulse amplitude modulation}
  \acro{PAPR}{Peak-to-Average Power Ratio}
  \acro{P2P}{Peer-to-Peer}
  \acro{PC}{Power Control}
  \acro{PCI}{Physical Cell ID}
  \acro{PDCCH}{physical downlink control channel}
  \acro{PDD}{penalty dual decomposition}
  \acro{PDF}{Probability Density Function}
  \acro{PER}{Packet Error Rate}
  \acro{PF}{Proportional Fair}
  \acro{P-GW}{Packet Data Network Gateway}
  \acro{PL}{Pathloss}
  \acro{RLT}{reformulation linearization technique}
  \acro{PRB}{Physical Resource Block}
  \acro{PROJ}{Projection-based Grouping}
  \acro{ProSe}{Proximity Services}
  \acro{PS}{\LU{P}{p}arameter \LU{S}{s}erver}
  \acro{PSO}{Particle Swarm Optimization}
  \acro{PUCCH}{physical uplink control channel}
  \acro{PZF}{Projected Zero-Forcing}
  \acro{QAM}{quadrature amplitude modulation}
  \acro{QoS}{quality of service}
  \acro{QPSK}{quadrature phase shift keying}
  \acro{QCQP}{quadratically constrained quadratic programming}
  \acro{RAISES}{Reallocation-based Assignment for Improved Spectral Efficiency and Satisfaction}
  \acro{RAN}{Radio Access Network}
  \acro{RA}{Resource Allocation}
  \acro{RAT}{Radio Access Technology}
  \acro{RATE}{Rate-based}
  \acro{RB}{resource block}
  \acro{RBG}{Resource Block Group}
  \acro{REF}{Reference Grouping}
  \acro{ReMAC}{Repetition for Multiple Access Computing}
  \acro{ReLU}{rectified linear unit}
  \acro{RF}{radio frequency}
  \acro{RLC}{Radio Link Control}
  \acro{RM}{Rate Maximization}
  \acro{RNC}{Radio Network Controller}
  \acro{RND}{Random Grouping}
  \acro{RRA}{Radio Resource Allocation}
  \acro{RRM}{\LU{R}{r}adio \LU{R}{r}esource \LU{M}{m}anagement}
  \acro{RSCP}{Received Signal Code Power}
  \acro{RSRP}{reference signal receive power}
  \acro{RSRQ}{Reference Signal Receive Quality}
  \acro{RR}{Round Robin}
  \acro{RRC}{Radio Resource Control}
  \acro{RSSI}{received signal strength indicator}
  \acro{RT}{Real Time}
  \acro{RU}{Resource Unit}
  \acro{RUNE}{RUdimentary Network Emulator}
  \acro{RV}{Random Variable}
  \acro{SAA}{Small Argument Approximation}
  \acro{SAC}{Session Admission Control}
  \acro{SCM}{Spatial Channel Model}
  \acro{SC-FDMA}{Single Carrier - Frequency Division Multiple Access}
  \acro{SD}{Soft Dropping}
  \acro{S-D}{Source-Destination}
  \acro{SDPC}{Soft Dropping Power Control}
  \acro{SDMA}{Space-Division Multiple Access}
  \acro{SDR}{semidefinite relaxation}
  \acro{SDP}{semidefinite programming}
  \acro{SER}{Symbol Error Rate}
  \acro{SES}{Simple Exponential Smoothing}
  \acro{S-GW}{Serving Gateway}
  \acro{SGD}{\LU{S}{s}tochastic \LU{G}{g}radient \LU{D}{d}escent}  
  \acro{SINR}{signal-to-interference-plus-noise ratio}
  \acro{SI}{self-interference}
  \acro{SIP}{Session Initiation Protocol}
  \acro{SISO}{\LU{S}{s}ingle \LU{I}{i}nput \LU{S}{s}ingle \LU{O}{o}utput}
  \acro{SIMO}{Single Input Multiple Output}
  \acro{SIR}{Signal to Interference Ratio}
  \acro{SLNR}{Signal-to-Leakage-plus-Noise Ratio}
  \acro{SMA}{Simple Moving Average}
  \acro{SNR}{\LU{S}{s}ignal-to-\LU{N}{n}oise \LU{R}{r}atio}
  \acro{SORA}{Satisfaction Oriented Resource Allocation}
  \acro{SORA-NRT}{Satisfaction-Oriented Resource Allocation for Non-Real Time Services}
  \acro{SORA-RT}{Satisfaction-Oriented Resource Allocation for Real Time Services}
  \acro{SPF}{Single-Carrier Proportional Fair}
  \acro{SRA}{Sequential Removal Algorithm}
  \acro{SRS}{sounding reference signal}
  \acro{SU-MIMO}{Single-User Multiple Input Multiple Output}
  \acro{SU}{Single-User}
  \acro{SVD}{Singular Value Decomposition}
  \acro{SVM}{\LU{S}{s}upport \LU{V}{v}ector \LU{M}{m}achine}
  \acro{TCP}{Transmission Control Protocol}
  \acro{TDD}{time division duplex}
  \acro{TDMA}{\LU{T}{t}ime \LU{D}{d}ivision \LU{M}{m}ultiple \LU{A}{a}ccess}
  \acro{TNFD}{three node full duplex}
  \acro{TETRA}{Terrestrial Trunked Radio}
  \acro{TP}{Transmit Power}
  \acro{TPC}{Transmit Power Control}
  \acro{TTI}{transmission time interval}
  \acro{TTR}{Time-To-Rendezvous}
  \acro{TSM}{Throughput-based Satisfaction Maximization}
  \acro{TU}{Typical Urban}
  \acro{UE}{\LU{U}{u}ser \LU{E}{e}quipment}
  \acro{UEPS}{Urgency and Efficiency-based Packet Scheduling}
  \acro{UL}{uplink}
  \acro{UMTS}{Universal Mobile Telecommunications System}
  \acro{URI}{Uniform Resource Identifier}
  \acro{URM}{Unconstrained Rate Maximization}
  \acro{VR}{Virtual Resource}
  \acro{VoIP}{Voice over IP}
  \acro{WAN}{Wireless Access Network}
  \acro{WCDMA}{Wideband Code Division Multiple Access}
  \acro{WF}{Water-filling}
  \acro{WiMAX}{Worldwide Interoperability for Microwave Access}
  \acro{WINNER}{Wireless World Initiative New Radio}
  \acro{WLAN}{Wireless Local Area Network}
  \acro{WMMSE}{weighted minimum mean square error}
  \acro{WMPF}{Weighted Multicarrier Proportional Fair}
  \acro{WPF}{Weighted Proportional Fair}
  \acro{WSN}{Wireless Sensor Network}
  \acro{WWW}{World Wide Web}
  \acro{XIXO}{(Single or Multiple) Input (Single or Multiple) Output}
  \acro{ZF}{Zero-Forcing}
  \acro{ZMCSCG}{Zero Mean Circularly Symmetric Complex Gaussian}
\end{acronym}

\title{ReMAC: Digital Multiple Access Computing \\
by Repeated Transmissions 
}

\graphicspath{{./Figures/}} 

\author{Xiaojing Yan,~\IEEEmembership{Member,~IEEE}, Saeed Razavikia,~\IEEEmembership{Member,~IEEE}, Carlo Fischione,~\IEEEmembership{Fellow,~IEEE}\\
 	\thanks{All the authors are with the School of Electrical Engineering and Computer Science KTH Royal Institute of Technology, Stockholm, Sweden (e-mail: xiay@kth.se, sraz@kth.se, carlofi@kth.se). C. Fischione is also with Digital Futures of KTH. }
    \thanks{S. Razavikia was supported by the Wallenberg AI, Autonomous Systems and Software Program (WASP), and  the Hans Werthén Foundation.}
 
    \thanks{A preliminary version of this work was presented in part at the IEEE International Conference on Communications, Denver, USA, June 2024, which appears in this manuscript as reference~\cite{yan2024novel}.}
}

\newtheorem{theorem}{Theorem}
\newtheorem{prop}{Proposition}
\newtheorem{lem}{Lemma}
\newtheorem{rem}{Remark}

\maketitle

\begin{abstract}
 In this paper, we consider the ChannelComp framework, where multiple transmitters aim to compute a function of their values at a common receiver while using digital modulations over a multiple access channel. ChannelComp provides a general framework for computation by designing digital constellations for over-the-air computation. Currently, ChannelComp uses a symbol-level encoding. However, encoding repeated transmissions of the same symbol and performing the function computation using the corresponding received sequence may significantly improve the computation performance and reduce the encoding complexity. In this paper, we propose a new scheme where each transmitter repeats the transmission of the same symbol over multiple time slots while encoding such repetitions and designing constellation diagrams to minimize computational errors. We formally model such a scheme by an optimization problem, whose solution jointly identifies the constellation diagram and the repetition code. We call the proposed scheme \ac{ReMAC}. To manage the computational complexity of the optimization, we divide it into two tractable subproblems. We verify the performance of \ac{ReMAC} by numerical experiments. The simulation results reveal that \ac{ReMAC} can reduce the computation error in noisy and fading channels by approximately up to $7.5$~dB compared to standard ChannelComp, particularly for product functions. 
\end{abstract}


\section{Introduction}

\acresetall

The next generation wireless technology, 6G, promises to give rise to numerous intelligent services via the edge network, utilizing distributed learning, data sensing, and analytics~\cite{wang2023road}. The enhanced capabilities of 6G will support massive connectivity, extremely low latency and
high energy efficiency, ultra-reliable transmission~\cite{mumtaz2021guest} for Internet of Things applications, spanning smart homes and cities to industrial automation and healthcare systems~\cite{wang2022overthe}.  In practice, tasks requiring functional computation result from distributed data, such as environment monitoring, autonomous control, and model updates, highlighting the demand for efficient communication protocols for data aggregation and computation~\cite{lim2020federated, xie2023networked}.  The current multi-access protocols are based on the orthogonal resource allocation among each device, which is resource-inefficient and unscalable~\cite{yin2006ofdma}. To address the challenge, a low latency non-orthogonal communication protocol called \ac{AirComp}~\cite{csahin2023survey,perez2024waveforms} is a promising joint communication and computation solution.

The primary concept of \ac{AirComp} is to leverage the waveform superposition property from the transmitted data of distributed nodes for function computation~\cite{chen2023over}.  This results in higher spectral efficiency and lower communication latency than conventional multiple-access schemes~\cite{goldenbaum2013harnessing}. Additionally, \ac{AirComp} exhibits scalability and adaptability to changing network conditions, making it well-suited for various network architectures, including multi-cell, hierarchical, and decentralized networks~\cite{wang2022overthe}.  Moreover, it has been proven that \ac{AirComp} helps support large-scale inference and learning tasks, particularly in communication-efficient federated learning~\cite{zhao2022broadband,amiri2020federated,hellstrom2022wireless}.

The previously discussed works predominantly assumed analog modulations, thereby limiting the practical applicability of \ac{AirComp} due to the rare support of such modulations in current wireless devices. Moreover, the lack of channel coding in analog communication hinders the requirement of highly reliable communication schemes. To overcome the \ac{AirComp} dependency on analog communications, a generalized communication-for-computation framework, termed ChannelComp~\cite{saeed2023ChannelComp}, has been introduced. This framework facilitates executing arbitrary finite functions over the \ac{MAC} using digital modulations. The ChannelComp designs the digital modulation in such a way as to make it possible to compute any function over-the-air. In fact, it extends the potential applications beyond the summation or nomographic operations supported by the traditional \ac{AirComp} method.

This paper presents a novel coding scheme for the ChannelComp framework that preserves ChannelComp's generality in computing functions over the air and provides a highly reliable communication scheme. We use transmission repetition together with {\color{black}coded repetition} and modulation design over multiple communication resources to perform computation.

\subsection{Literature Review}

The problem of computing linear functions over a multiple-access channel has been first studied in~\cite{nazer2007computation,nazer2011compute}, where a computation coding scheme has been developed for distributed computation. In~\cite{goldenbaum2014nomographic,goldenbaum2015achievable}, the computation capacity of \ac{AirComp} has been extended to a class of functions known as nomographic functions, which possess a structure that leverages the interference property of \ac{AirComp}. \ac{AirComp} has gained popularity in different disciplines, such as signal processing~\cite{zhu2018mimo,razavikia2022blind,daei2024timely}, power management~\cite{zhai2020power}, and privacy preservation~\cite{massny2023secure}, thanks to its simplicity in scaling and reducing the complexity of encoding and decoding operations~\cite{jha2022fundamental}. Nevertheless, since the signals are directly transmitted over wireless channels without encoding, \ac{AirComp} uses analog modulations and thus lacks robust mechanisms to counteract noise, which leads to increased computation errors and decreased reliability~\cite{csahin2022over,yao2024digital}. Moreover, \ac{AirComp} is incompatible with the predominantly digital infrastructure of current wireless systems.

To address these challenges, there is rising interest in designing digital \ac{AirComp} versions that leverage advanced source and channel coding capabilities~\cite{hellstrom2022wireless,razavikia2023blind}. One of the earliest attempts involves implementing signSGD for federated learning problems~\cite{zhu2020one}. Additionally, variants such as the phase asynchronous OFDM-based OBDA~\cite{you2023broadband} and non-coherent communication solutions for AirComp~\cite{csahin2022over,sahin2024over} have been proposed. Furthermore, targeting federated edge learning, in~\cite{qiao2024massive}, vector quantization has been employed to reduce uplink communication overhead using shared quantization and modulation codebooks. These studies mainly focus on particular machine learning training processes (e.g., signSGD~\cite{bernstein2018signsgd}), working with few transmitters or assuming that the number of transmitting devices is much smaller than the codebook size.

Recent studies have introduced a digital channel computing framework called ChannelComp~\cite{razavikia2023computing,saeed2023ChannelComp,razavikia2024sumcomp}, providing a general framework for performing function computation over the \ac{MAC}. ChannelComp enhances computation accuracy and system reliability by designing digital modulation to make the computation of any desired function feasible over the \ac{MAC}. Inspired by~\cite{razavikia2024sumcomp}, authors in~\cite{liu2024digital} have presented a framework featuring digital modulation of each data value integrated with bit-slicing techniques to allocate its bits to multiple symbols, thereby increasing digital \ac{AirComp} reliability. Furthermore,~\cite{xie2023joint} has proposed a joint design of channel coding and digital modulation for digital \ac{AirComp} transmission, reducing complexity by transforming digital \ac{AirComp} into an ordinary point-to-point system and adopting standard channel coding schemes such as non-binary low-density parity-check code.

However, these codes are often restricted to digital modulation formats, such as \ac{QAM}, and to computing nomographic functions~\cite{razavikia2024sumcomp}. Consequently, they may not generalize well to more complex modulation schemes or arbitrary classes of functions as proposed in ChannelComp~\cite{saeed2023ChannelComp}.

\begin{table}[t]
\caption{ Reference list of commonly used variables in this paper.}
\begin{center}
\begin{tabular}{|c|c|}
\hline
\textbf{Variable} & \textbf{Definition} \\
\hline
$f$ & Desired function \\
$K$ & Number of nodes \\
$Q$ & Number of quantization levels \\
$L$ & Number of time slots \\
$x_k$ & Input value of node $k$ \\
$\tilde{c}_{k,\ell}$ & Channel code of node $k$ at time slot $\ell$ \\
$\Vec{x}_k$ & Modulated signal of node $k$ \\
$\mathcal{E}_k(\cdot)$ & Modulation encoder of node $k$ \\
$\mathcal{C}_k(\cdot)$ & Channel encoder of node $k$ \\
$\mathcal{T}(\cdot)$ & Tabular mapping \\
$h_{k,\ell}$ & Channel coefficient between node $k$ and CP at time slot $\ell$ \\
$p_{k,\ell}$ & Transmit power of node $k$ at time slot $\ell$ \\
$\Vec{z}_{\ell}$ & Additive white Gaussian noise at time slot $\ell$ \\
$\vec{s}_{k,\ell}$ & Transmitted signal by node $k$ at time slot $\ell$ \\
$\vec{y}_{\ell}$ & Received signal by the CP at time slot $\ell$ \\
\hline
\multicolumn{2}{p{0.9\linewidth}}{ }
\label{notation table}
\end{tabular}
\end{center}
\vspace{-10pt}
\end{table}

\input{Figures/Fig-system-new}

\subsection{Contribution}

This paper proposes a novel and general scheme for computing functions over the MAC, termed  \ac{ReMAC}. Inspired by ChannelComp, \ac{ReMAC} jointly designs digital modulations and channel codes of repeated symbol sequences over multiple time slots. More specifically, for the transmission, we consider modulation and channel encoders, where the modulation encoder maps the input value to the constellation diagram of the modulation, and the channel encoder selects the constellation points over multiple time slots. We provide the conditions to avoid destructive overlaps among the constellation points, which allows us to perform a valid function computation at the receiver. Then, we jointly design modulation and channel encoders through an optimization problem that ensures a valid function computation across the consecutive time slots. Due to the non-convex and NP-hard nature, the problem is relaxed, and an approximate solution is developed by alternating minimization to obtain modulation and channel encoders. We provide the convergence analysis and the optimality gap for the proposed approximate solution method. Finally, we evaluate the reliability performance of  \ac{ReMAC} via comprehensive simulation results.  Particularly, our contributions are  listed below:

\begin{itemize}
    \item \textbf{Reliable communication}: We introduce \ac{ReMAC}, a novel approach that combines repeated transmission scheme that extends the ChannelComp framework. By selectively transmitting repeated modulated symbols over multiple time slots, \ac{ReMAC} aims to enable computation over the \ac{MAC} by digital communications. This method leverages {\color{black}coded repetition} to introduce redundancy into the transmissions, thereby reducing computation errors.
    
    \item \textbf{Joint modulation and {\color{black}coded repetition} design}: We establish the necessary conditions for valid function computation, resulting in a non-convex optimization problem for designing the constellation points of digital modulation and their allocation across multiple time slots. Given the NP-hard nature of this problem, we propose a strategy to manage its complexity. Specifically, we decompose the problem into two subproblems and apply convex approximations. Then, the resulting problem is solved using an alternating minimization approach.
    
    \item \textbf{Theoretical analysis}: We present the convergence rate for the proposed alternating minimization approach of \ac{ReMAC}. Our analysis demonstrates that the optimality gap between the obtained approximate solution and the true optimal solution depends on the number of constraints in the original optimization problem and the initial values of the optimization variables. Additionally, we prove that \ac{ReMAC} can achieve the first-order stationary point of the primal optimization problem for the joint design of modulation and {\color{black}coded repetition}.
    
    \item \textbf{Numerical experiments}: We provide comprehensive numerical experiments to validate the computation performance of \ac{ReMAC} over noisy and fading channels. Simulation results reveal that, even in the presence of fading, \ac{ReMAC} can effectively reduce the computation errors by up to approximately $7.5$ dB compared to the ChannelComp framework, particularly for the product function.  
\end{itemize}

\subsection{Organization of the Paper}

The paper is organized as follows: In Section~\ref{sec:Systemmodel}, we provide an overview of the system model and explain the overlaps of constellation points. Then, we propose the problem of jointly designing and selecting digital modulation formats in Section~\ref{subsec:ReMAC design}. Moreover, we provide the theoretical guarantees regarding the proposed alternating algorithm in Subsection~\ref{sec:Theory}. In Section~\ref{sec:Num}, we present the results of numerical experiments, comparing our proposed \ac{ReMAC} with ChannelComp and digital \ac{AirComp}. Finally, in Section~\ref{sec:conclusion}, we conclude the paper and outline potential directions for future research.

\section{System Model}\label{sec:Systemmodel}

\subsection{Signal Model}

\input{Figures/Fig_example}

Consider a network with $K$ single antenna nodes and a server called \ac{CP}, where all the nodes communicate with the \ac{CP} over a shared channel. Let $x_k \in  \mathbb{F}_Q$ be the value owned by node $k$, where $\mathbb{F}_Q$ is a finite field with $Q$ unique values, referred to as the quantization level. Note that we consider $x_k$ to be quantized using $b$-bit, in which $b = \log{(Q)}$, and all the values are stored and processed in digital format. Given $x_1,\ldots,x_K$ at the nodes, the network's main goal is to compute function $f(x_1, \ldots, x_K): \mathbb{F}_Q 
 \mapsto \mathbb{F}_{Q'}$ over the MAC with $L$ time slots.  To this end, the value ${x}_k$ is converted into a digital modulation signal in the complex domain using a modulation encoder $\mathcal{E}_k(\cdot): \mathbb{F}_Q \mapsto \mathbb{C}$, i.e., $\vec{x}_k = \mathcal{E}_k({x}_k) \in \mathbb{C}$. In parallel, the quantized values are selectively retransmitted across $L$ consecutive time slots, each processed by a channel encoder denoted as $\mathcal{C}: \mathbb{F}_Q \mapsto \mathbb{F}_2^{L}$. The generated $L$-length binary sequence $\tilde{\mathbf{c}}_{k} = \mathcal{C}_k({x}_k) \in \{0,1\}^{L}$ indicates the specific time slots to which ${x}_k$ is allocated. Afterward, all $K$ nodes simultaneously transmit the modulated symbols $\vec{x}_k$ within the assigned time slots, and the \ac{CP} receives the superimposed signal as each time slot by.  
\begin{align}
\label{eq:receivedsignal}
 \vec{y}_{\ell} = \sum\nolimits_{k=1}^{K} h_{k,\ell}p_{k,\ell}\vec{x}_k\cdot \tilde{c}_{k,\ell}+\vec{z}_{\ell},\quad \forall~\ell \in [L],   
\end{align}
where $\tilde{c}_{k,\ell}$ denotes the $\ell$-th entry of vector $\tilde{{\mathbf{c}}}_{k}$. Moreover, $h_{k,\ell}$ denotes the channel coefficient between node $k$ and \ac{CP} at time slot $\ell$, and $\vec{z}_{\ell}$ represents the additive white Gaussian noise at time slot $\ell$ \footnote{Note that the channels between the nodes and the \ac{CP} are assumed to be block-fading and remain unchanged within every time slot. The signals transmitted over-the-air experience additive white Gaussian noise.}. With perfect \ac{CSI} and employing an optimal power control policy~\cite{cao2020optimal}, the transmit power of node $k$ at time slot $\ell$ is determined as the inverse of the channel coefficient, denoted as $p_{k,\ell}=h_{k,\ell}^*/|h_{k,\ell}|^2$. Consequently, Eq.~\eqref{eq:receivedsignal} can be rewritten as follows.
\begin{align}
\label{eq:nofading}
\vec{y}_{\ell} = \sum\nolimits_{k=1}^{K} \vec{s}_{k,\ell} + \vec{\tilde{z}}_{\ell}, \quad \forall~\ell \in [L],
\end{align}
where $\vec{\tilde{z}}_{\ell}\sim \mathcal{N}(0,\sigma_{z}^2)$ involves the error that results from the imperfect channel compensation together with the receiving noise~\cite{cao2019optimal}. Also, $\vec{s}_{k,\ell}:=  \vec{x}_k \cdot \tilde{c}_{k,\ell}$ represents the digitally modulated symbol transmitted at time slot $\ell$ by node $k$.

Note that, due to the digital modulation formats, each transmitted signal $\vec{s}_{k,\ell}$ is constrained to a finite set of constellation points; therefore, the received signal $\vec{y}_{\ell}$ yields a finite constellation diagram (ignoring the effect of the noise $\vec{\tilde{z}}_{\ell}$).
To compute the desired function $f$ from the received sequence of $\bm{y}:= [\vec{y}_{1}, \ldots, \vec{y}_{L}]$, the \ac{CP} can apply a tabular mapping $\mathcal{T}$ to the sequence of constellation points to obtain the function output value, as long as such received sequence can be uniquely associated to the function's output. The complete process of \ac{ReMAC} is depicted in Figure~\ref{fig:digital_aircomp_new}.

{\color{black}
\begin{rem}
Note that ReMAC assumes a phase-aligned precoded transmission. However, physical system imperfections, such as carrier frequency offset, synchronization delays, and hardware-induced phase noise, may introduce precoding errors~\cite{tan2018mobile,perez2024waveforms}. To mitigate these errors, techniques such as pre-equalization can be incorporated to correct waveform misalignments before transmission, as studied in~\cite{guo2021over}.
\end{rem}}

In the next subsection, we explain in detail the overlaps of constellation points and present how to avoid them in designing the modulation diagram and the channel codes. 

\subsection{Overlaps of Constellation Points }\label{sec:ProblemFormualtion}

In this subsection, we present the necessary and sufficient conditions for computing the function over the \ac{MAC} without any destructive overlaps among constellation points of superimposed transmitted signals. Considering a noiseless \ac{MAC}, let $x_1^{(i)}, \ldots, x_K^{(i)}$ and $x_1^{(j)}, \ldots, x_K^{(j)}$ be two sets of input values, with their corresponding output values of the function $f(x_1, \ldots, x_K)$ being $f^{(i)}$ and $f^{(j)}$, respectively. The aggregated constellation points at time slot $\ell$ are then given by $\Vec{v}_{\ell}^{(i)}:= \sum_{k}\vec{s}_{k,\ell}^{(i)}$
and $\Vec{v}_{\ell}^{(j)}:= \sum_{k}\vec{s}_{k,\ell}^{(j)}$. To ensure the \ac{CP} can accurately compute the function output $f^{(i)}$ and $f^{(j)}$ from the corresponding sequences of the induced points $\bm{v}^{(i)}:=[\Vec{v}_{1}^{(i)},\ldots,\Vec{v}_{L}^{(i)}]$ and $\bm{v}^{(j)}:=[\Vec{v}_{1}^{(j)},\ldots,\Vec{v}_{L}^{(j)}]$,  these sequences need to be distinguishable. Indeed, for two distinct output values $f^{(i)}$ and $f^{(j)}$, we need two distinct sequences of $\bm{v}^{(i)}$ and $\bm{v}^{(j)}$, i.e., $\bm{v}^{(i)}\neq \bm{v}^{(j)}$.  Mathematically, all we need to have a valid computation over the \ac{MAC} is to satisfy the following constraints~\cite{razavikia2023computing}:
\begin{align}
    \label{eq:CompCond}
    {\rm if }~f^{(i)}\neq f^{(j)}~\Rightarrow~\bm{v}^{(i)} \neq \bm{v}^{(j)},\quad \forall (i,j)\in [M]^2,
\end{align}
where $M$ is the cardinality of the range of function $f$, and $[M]^2 = [M] \times [M]$ denotes the Cartesian product. 
The main idea here is to jointly design $\mathcal{E}_k$ and $\mathcal{C}_k$ under the constraint in \eqref{eq:CompCond}, allowing the tabular mapping $\mathcal{T}$ to uniquely associate the output value $f^{(i)}$ with $\bm{v}^{(i)}$ and compute the exact output value. More precisely,  the \ac{CP} uses a set of constellation points ${\bm{v}}^{(i)} = [\vec{v}^{(i)}_{1}, \ldots, \vec{v}^{(i)}_{L}]^{\mathsf{T}} \in \mathbb{C}^{L \times 1}$ to obtain the function output $f^{(i)}$. Consequently, the goal becomes to identify the corresponding points of ${\bm{v}}^{(i)}$ when $\bm{y}$ is received, which can be achieved by using the maximum likelihood estimator for $\bm{y}$. Following similar steps as in~\cite{razavikia2023computing}, one can show that the maximum likelihood estimator reduced to the following problem.
\begin{align}
\label{eq:MLE_esitmator}
\hat{f}^{(i)}=\arg \min_{i}||\bm{y} - \bm{v}^{(i)}||_2^2.
\end{align}

Then, the constellation diagram, which consists of all possible constellation points $\{\vec{v}_{\ell}^{(1)},\ldots,\vec{v}_{\ell}^{(M)}\}$ along with their corresponding Voronoi cells $\{\mathcal{V}_{1,\ell},\ldots,\mathcal{V}_{M,\ell}\}$, are generated by the decoder. The Voronoi cells contain the points in the signal space that are closest to a specific constellation point. The desired function $\hat{f}$  is computed over all $L$ time slots by $\hat{f}=\sum_{j=1}^M\mathcal{T}_{j}(\vec{\bm{v}})$, where $\mathcal{T}_{j}(\cdot)$ is an indicator function: 
\begin{align}
\mathcal{T}_{j}(\vec{\bm{v}}) := \begin{cases}
    \hat{f}^{(j)}, & \quad \text{if } \vec{v}_{\ell} \in \mathcal{V}_{j,\ell},\text{ } \forall \ell \in [L], \\
0, & \quad \textrm{otherwise}. 
\end{cases}    
\end{align}

We merge the Voronoi cells of overlapping points into a single cell. Specifically, for points whose corresponding constraints in~\eqref{eq:CompCond} are not satisfied, we assign one Voronoi cell. Consequently, we must assign a single output value for multiple function outputs. For instance, we can use the average output values corresponding to the merged Voronoi cells as the final output value~\cite{saeed2023ChannelComp}.

\begin{rem}
    Note that one can use a maximum posterior estimator for the decoding parts instead of the maximum likelihood estimator. The overall decoding strategy using tabular map $\mathcal{T}$ remains the same, and only in finding the boundary of  Voronoi cells using  \eqref{eq:MLE_esitmator} needs to be changed. 
\end{rem}

\subsection{How Can Repetition Resolve the Overlaps?}
 
In this subsection, we clarify the role of repetition in resolving the overlapping constellation points that may otherwise occur at the CP. In what follows, we provide a simple example to give better insight into the performance of the \ac{ReMAC}. Indeed, consider the case of a network with $K=4$ nodes aimed at computing the product function $f(x_1,x_2,x_3,x_4) = x_1x_2x_3x_4$ over the noiseless \ac{MAC}. Each node has input values $x_k \in \{1, 2, 3, 4\}$ and uses a simple quadrature phase shift keying (QPSK) modulation for the transmission. These input values are then encoded as $\vec{x}_k \in \{1, -1, i, -i\}$.

In~\cite{saeed2023ChannelComp}, it was shown that computing the product function requires either changing the constellation points from QPSK to a new modulation diagram or increasing the order of modulation to avoid destructive overlaps. Instead, \ac{ReMAC} proposes transmitting the modulated symbol repeatedly over multiple time slots, i.e., $L=2$, and allocating a subset of constellation points to each time slot to resolve destructive overlaps. In the simple illustrative example of this subsection, the ReMAC scheme employs a specialized {\color{black} repetition} encoder to transmit selectively each modulated symbol as follows:
\begin{equation} \nonumber
\tilde{c}_{k,0} = \left\{
\begin{aligned}
1, & ~ \text{if } {x}_k \in \{1,3\} \\
0, & ~ \text{if } {x}_k \in \{2,4\}
\end{aligned}
\right.  \\
,\tilde{c}_{k,1} = \left\{
\begin{aligned}
0, & ~ \text{if } {x}_k \in \{1,3\} \\
1, & ~ \text{if } {x}_k \in \{2,4\}
\end{aligned}
\right..
\end{equation}

Figures~\ref{fig:conflicts} and~\ref{fig:error-free} show examples with two specific sets of input values, $x_k \in (1,1,2,2)$ and $x_k\in (1,2,3,4)$.  From Figure~\ref{fig:conflicts}, we observe that the resultant constellation diagram overlaps at the constellation point $0$, corresponding to two different output values, $\{4, 24\}$. For the ChannelComp framework, since using QPSK results in overlaps of the constellation points, the overlaps are unresolved no matter how often we repeat the transmission.  However, \ac{ReMAC} applies {\color{black} repetition encoder} $\mathcal{C}(\cdot)$ to the transmitted constellation points, making the induced sequence from the received signals $\vec{y}_{\ell} = \sum_{k=1}^{K} \vec{x}_k \cdot \tilde{c}_{k,\ell}$ distinct across time slots. Consequently, the corresponding function outputs can be computed using a tabular map $\mathcal{T}$ by simply referencing this sequence.

\begin{rem}
Note that the \ac{ReMAC} assumes that the number of time slots $L$ satisfies $1 < L < \min\{Q, K\}$, where $K$ is the number of nodes, and $Q$ is the number of constellation points. Specifically, we highlight three noteworthy cases. 
First, when $L = K < Q$, each node is allocated a distinct time slot for transmission, aligning with the orthogonal resource allocation. This allows the CP to decode each transmitted symbol individually and perform the required computations afterward.
Second, when $L = Q < K$, each time slot is assigned a specific constellation point. This configuration is particularly effective for symmetric functions because it enables the CP to determine the number of nodes transmitting each constellation point. Consequently, the CP can reconstruct the output of the symmetric function.
Finally, in the special case where $L=1$, our \ac{ReMAC} simplifies to ChannelComp~\cite{razavikia2023computing}, requiring all the designed constellation points to be transmitted within a single time slot for computing the desired functions.
\end{rem}

{\color{black} The ReMAC does not lose spectral efficiency compared to ChannelComp and the other state-of-the-art methods. Indeed, the ReMAC is spectral efficiency as the ChannelComp framework, where computing functions $f$ with $K$ nodes only require a single frequency resource. The ReMAC uses time diversity to enhance computational reliability.  While repetition effectively resolves overlapping constellation points and provides computational reliability,  it leads to a higher latency communication protocol. For instance, when \( L < K \), the repetition strategy achieves superior latency efficiency compared to the traditional \ac{TDMA}, where the transmission of each node occurs over distinct time slots. For the special case where \( L = K \), the latency of the ReMAC becomes the same as the \ac{TDMA}. Thus, the choice of strategy depends on the specific scenario, requiring a balance between computational reliability and latency.}

Considering the distortion introduced by noise term $\vec{\tilde{z}}_{\ell}$, the output of tabular map $\mathcal{T}$  may result in computation errors. In the next section, we jointly design encoders and decoders for \ac{ReMAC} to enable the reliable computation of the desired function $f$ over the \ac{MAC}.

\section{The proposed \ac{ReMAC}} \label{subsec:ReMAC design}

To obtain a proper set of the modulation encoder $\mathcal{E}_k$ and {\color{black}repetition} encoder $\mathcal{C}_k$ for all nodes $k \in [K]$, we need to jointly design the constellation diagram and the repetition coding strategy so that the corresponding sequences of the induced points $\bm{v}$ satisfy \eqref{eq:CompCond}. This joint design aims to achieve robust and power-efficient computation over the \ac{MAC} while avoiding destructive overlaps. Consequently, the tabular mapping $\mathcal{T}$ can uniquely map the constellation points of $\Vec{y}_{1}, \ldots, \Vec{y}_{L}$ to the corresponding outputs of the function $f$. 
More precisely, let the modulation vector $\bm{x}_k := [\vec{x}_k^{(1)}, \ldots, \vec{x}_k^{(Q)}]^{\mathsf{T}}\in \mathbb{C}^{Q \times 1}$  contain all $Q$ possible constellation points created by the modulation encoder $\mathcal{E}_k(\cdot)$. Then, by concatenating all the $K$ modulation vectors, we define the modulation vector  as $\bm{x}:= [\bm{x}_1^{\mathsf{T}}, \ldots, \bm{x}_K^{\mathsf{T}}]^{\mathsf{T}} \in \mathbb{C}^{N \times 1}$, where $N:= Q\times K$. Similarly, we define the symbol allocations matrix of node $k$ as $\bm{C}_{k} \in \{0,1\}^{Q \times L}$, where the element $[\bm{C}_{k}]_{(\ell,q)} := c_{k,\ell}^{(q)} \in \{0,1\}$ indicates whether the modulated signal $ x_k^{(q)} $ is transmitted over time slot $\ell$. We further define the {\color{black}repetition coding matrix} as $\bm{C} := [\bm{C}_1^{\mathsf{T}}, \ldots, \bm{C}_K^{\mathsf{T}}]^{\mathsf{T}} \in \{0,1\}^{N \times L}$ to denote the code for the repetition.
Therefore, the resultant constellation point sequence $\bm{v}^{(i)}$ corresponding to output value $f^{(i)}$ can be formulated in matrix notation as follows:
\begin{equation}
\bm{v}^{(i)} = \bm{a}_i^{\mathsf{T}} (\bm{x} \otimes \mathds{1}_{L}^{\mathsf{T}}) \odot \bm{C} \in  \mathbb{C}^{1 \times L}, \quad \forall i \in [M],
\end{equation}
where $\mathds{1}_{L}$ denotes an $L$-length column vector of ones, and $\bm{a}_i^{\mathsf{T}}$ is a binary vector that selects the support of the input values associated with $f^{(i)}$. To satisfy the computation constraint in \eqref{eq:CompCond} and to incorporate the noise effect, we can replace such a non-smooth constraint for a pair $f^{(i)}$ and $f^{(j)}$ with their corresponding  vectors $\bm{v}^{(i)}$ and $\bm{v}^{(j)}$ by the following constraint:
\begin{align}
    \label{eq:CondMain}
    \| \bm{v}^{(i)} - \bm{v}^{(j)}\|_{2}^2 \geq  \sigma_z^2|f^{(i)} - f^{(j)}|, \text{ } \forall (i,j) \in [M]^2,
\end{align}
where $\|\cdot\|_2$ denotes the $\ell_2$ norm. The inequality in~\eqref{eq:CondMain} ensures that resultant constellation points are distinguishable for two different $f^{(i)}$ and $f^{(j)}$. Without loss of generality, we assume that  $\|\bm{x}\|_2^2 \leq 1$. Then, to obtain a robust and power-efficient coding scheme over the MAC, we pose the following optimization problem.
\begin{subequations}
\label{eq:original problem 0}
\begin{align}
\nonumber
 \mathcal{P}_{0} := \underset{\bm{x},\bm{c}_{\ell}}{ \text{min}} ~& \sum_{\ell =1}^L\|\bm{c}_{\ell}\|_1, \\ \label{eq:smooth original problem}
{\rm s.t.}~& \sum\nolimits_{\ell=1}^L |(\bm{a}_i-\bm{a}_j)^{\mathsf{T}}(\bm{x}\odot\bm{c}_{\ell})|^2 \geq \Delta f_{i,j},  \\ 
& \forall (i,j) \in [M]^2,  \quad \|\bm{x}\|_2^2 \leq 1, \\
& \bm{c}_{\ell} \in \{0,1\}^{N \times 1}, \quad \forall \ell \in [L],
\end{align}
\end{subequations}
where $\Delta f_{i,j}:=\sigma_z^2 |f^{(i)}-f^{(j)}|$.  Problem $\mathcal{P}_0$ jointly optimizes the modulation vector $\bm{x}$ and channel vectors $\bm{c}_{\ell}$ while guaranteeing that the computation constraints hold. By solving Problem $\mathcal{P}_0$, we acquire the optimum $\bm{x}^*$ and $\bm{c}_{\ell}^*$s for $\ell \in [L]$. Thereafter, we obtain the modulation encoders $\mathcal{E}_1(\cdot), \ldots, \mathcal{E}_K(\cdot)$  and {\color{black} repetition} encoders $\mathcal{C}_1(\cdot), \ldots, \mathcal{C}_K(\cdot)$, respectively, that map the input values to the resultant modulation vector and repetition coding matrix. However, due to the nonconvex quadratic constraints in  \eqref{eq:smooth original problem}, Problem $\mathcal{P}_0$ is NP-hard~\cite{saeed2023ChannelComp}. To circumvent the complexity of Problem $\mathcal{P}_0$, we provide an efficient algorithm to compute an approximate solution in the next subsection. 
\subsection{Joint Design of Modulation and {\color{black}Repetition Coding}}\label{sec:MethodDesign}
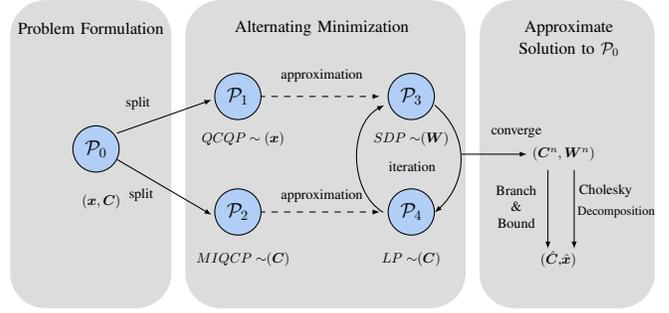
\begin{figure}
\scalebox{0.53}{
\centering

\tikzset{every picture/.style={line width=0.75pt}}        

\begin{tikzpicture}[x=0.75pt,y=0.75pt,yscale=-1,xscale=1]

\draw[draw opacity=0][fill={rgb, 255:red, 220; green, 220; blue, 220 } , rounded corners=25pt] (225pt, 800pt) rectangle (425pt, 580pt) {};

\draw[draw opacity=0][fill={rgb, 255:red, 220; green, 220; blue, 220 } , rounded corners=25pt] (100pt, 800pt) rectangle (215pt, 580pt) {};

\draw[draw opacity=0][fill={rgb, 255:red, 220; green, 220; blue, 220 } , rounded corners=25pt] (435pt, 800pt) rectangle (560pt, 580pt) {};

\draw[-latex]    (537.29,873.29) .. controls (570.61,900.73) and (569.35,946.41) .. (537.29,975.52) ;

\draw[-latex]   (488.29,975.29) .. controls (455.95,956.67) and (454.48,891.95) .. (489.25,873.36) ;

\draw[-latex]    (235,900) -- (325,868.32) ;

\draw[-latex]    (235,925) -- (325,977.74) ;

\draw[-latex]  [dash pattern={on 4.5pt off 4.5pt}]  (370,865) -- (485,865) ;

\draw[-latex]  [dash pattern={on 4.5pt off 4.5pt}]  (370,975) -- (485,975) ;

\draw[-latex]     (563,920) -- (625,920) ;

\draw[-latex]    (670,935) -- (670,1010) ;

\draw[-latex]    (645,935) -- (645,1010) ;

\draw  [fill={rgb, 255:red, 173; green, 205; blue, 250 }  ,fill opacity=0.9 ]  (215, 915) circle (22)   ;
\draw (215,915) node  [font=\Large]  {$\mathcal{P}_{0}$};

\draw  [fill={rgb, 255:red, 173; green, 205; blue, 250 }  ,fill opacity=0.9 ]  (350, 865) circle (22)   ;
\draw (350, 865) node  [font=\Large]  {$\mathcal{P}_{1}$};

\draw  [fill={rgb, 255:red, 173; green, 205; blue, 250 }  ,fill opacity=0.9 ]  (350, 975) circle (22)   ;
\draw (350, 975) node  [font=\Large]  {$\mathcal{P}_{2}$};

\draw  [fill={rgb, 255:red, 173; green, 205; blue, 250 }  ,fill opacity=0.9 ]  (515, 865) circle (22)   ;
\draw (515, 865) node  [font=\Large]  {$\mathcal{P}_{3}$};

\draw  [fill={rgb, 255:red, 173; green, 205; blue, 250 }  ,fill opacity=0.9 ]  (515, 975) circle (22)   ;
\draw (515, 975) node  [font=\Large]  {$\mathcal{P}_{4}$};

\draw (210,800) node  [font=\large]  {Problem  Formulation};

\draw (430,800) node  [font=\large]  {Alternating  Minimization};

\draw (700,955) node     {Cholesky};
\draw (710,975) node   [font=\small]  {Decomposition};
\draw (665,800) node  [font=\large]  {Approximate };
\draw (665,820) node  [font=\large]  {Solution to $\mathcal{P}_0$};

\draw (655,1020) node   {$ (\hat{\bm{C}}$,$ \hat{\bm{x}})$};

\draw (660,920) node   {$(\boldsymbol{C}^{n} ,\boldsymbol{W}^{n})$};

\draw (220,965) node   {$(\bm{x},\bm{C})$};
\draw (355,905) node    {$QCQP\sim (\bm{x})$};
\draw (355,1020) node    {${MIQCP\sim }(\bm{C})$};
\draw (515,905) node    {${SDP\sim}(\bm{W})$};
\draw (515,1020) node    {${LP\sim}(\bm{C})$};
\draw (430,960) node  {approximation};

\draw (430,845) node  {approximation};

\draw (515.94,931.5) node  {iteration};
\draw (254.94,871.5) node  {split};
\draw (257.94,958.5) node  {split};
\draw (615,900) node  {converge};
\draw (615,970) node  [color={rgb, 255:red, 0; green, 0; blue, 0 }  ,opacity=1 ] [align=left] {\begin{minipage}[lt]{39.55pt}\setlength\topsep{0pt}
\begin{center}
Branch \\ \& \\Bound
\end{center}

\end{minipage}};

\end{tikzpicture}
}
\caption{Flow diagram of the proposed alternating minimization algorithm, formally given in Algorithm~\ref{Alg:HIT} later in the paper.  
}
\label{fig:flowdigram}
\end{figure}

To approximately solve Problem $\mathcal{P}_{0}$, we alternately optimize the objective with respect to the variables $\bm{x}$ and $\bm{c}_{\ell}$s, resulting in two subproblems. Then, we relax each subproblem separately to obtain their optimum solutions. In Figure~\ref{fig:flowdigram}, we depict the flow chart diagram of the overall procedure. Specifically, for a given $\bm{c}_{\ell}^{n-1}$ at iteration $n$,  the optimization Problem $\mathcal{P}_0$ over $\bm{x}$ turns into the following feasibility problem. 
\begin{subequations}
\label{eq:CVX}
\begin{align}
    \nonumber
    \mathcal{P}_{1} & =  {\rm find} ~~\bm{x}, \\
    \text{s.t.} &\ \sum\nolimits_{\ell=1}^L |(\bm{a}_i-\bm{a}_j)^\top(\bm{x}\odot\bm{c}_{\ell}^{n-1})|^2 \geq \Delta f_{i,j}, \\
    & \quad \|\bm{x}\|_2^2 \leq 1.
\end{align}
\end{subequations}

To optimize over the variables $\bm{c}_{\ell}$s, we remove the power constraints imposed on $\bm{x}$. As a result, Problem $\mathcal{P}_0$ with respect to $\bm{c}_{\ell}$s becomes
\begin{subequations}
\label{eq:MBO}
\begin{align}
    \nonumber
    \mathcal{P}_{2} & = \min_{\bm{c}_{\ell}}  \sum\nolimits_{\ell=1}^L|| \bm{c}_{\ell}||_1, \\
    \text{s.t.} & \sum\nolimits_{\ell=1}^L |(\bm{a}_i-\bm{a}_j)^\top(\bm{x}^{n}\odot\bm{c}_{\ell})|^2 \geq \Delta f_{i,j}, \\
    & \bm{c}_{\ell} \in \{0,1\}^{N}, \quad \forall \ell \in [L].
\end{align}
\end{subequations}

Both Problems $\mathcal{P}_1$ and $\mathcal{P}_2$ are NP-hard and challenging to solve. In what follows, we provide the relaxation techniques and approximation solutions for both problems. 

\subsubsection{Optimize for the Modulation Vector}

The optimization Problem $\mathcal{P}_1$ in \eqref{eq:CVX} is a \ac{QCQP} problem, which is non-convex and NP-hard~\cite{sidiropoulos2006transmit}. To address the non-convexity, we employ the lifting trick~\cite{vandenberghe1996semidefinite}, which transforms the non-convexity into a rank constraint. By relaxing the problem and dropping the rank constraint, we obtain a convex optimization problem. In particular, we rewrite Problem $\mathcal{P}_1$ in terms of a new lifted matrix variable $\bm{W}:= \bm{x} \bm{x}^{\mathsf{H}}$ as 
\begin{subequations}
\label{eq:LiftSDP}
\begin{align}
\nonumber
\mathcal{P}_{1} := &  {\rm find} ~~ \bm{W},      \\ \label{eq:lifting CVX}
{\rm s.t.} \quad & {\rm Tr}(\bm{W} \cdot \bm{B}_{i,j}^{n-1}) \geq \Delta f_{i,j},\\
 & \bm{W} \succeq \bm{0}, \quad {\rm rank}(\bm{W})=1,  \label{eq:lift-rank} \\
& \text{Tr}(\mathbf{W}) \leq 1,
\end{align}
\end{subequations}
where $$\bm{B}_{i,j}^{n-1}=\sum\nolimits_{\ell=1}^L((\bm{a}_i-\bm{a}_j)\odot\bm{c}_{\ell}^{n-1})((\bm{a}_i-\bm{a}_j)\odot\bm{c}_{\ell}^{n-1})^\top.$$ To relax the problem, we drop the rank-one constraint from \eqref{eq:lift-rank}, which yields the following.   
\begin{subequations}
\label{eq:LiftSDPconvex}
\begin{align}
\nonumber
\mathcal{P}_{1} \approx \mathcal{P}_{3} := &  {\rm find} ~\bm{W},  \\ \label{eq:lifting CVX-relaxed}
{\rm s.t.} \quad & {\rm Tr}(\bm{W} \cdot \bm{B}_{i,j}^{n-1}) \geq \Delta f_{i,j},\\
 & \bm{W} \succeq \bm{0}, \quad \text{Tr}(\mathbf{W}) \leq 1.  \label{eq:lift-rank-relaxed} 
\end{align}
\end{subequations}

Problem $\mathcal{P}_3$ is convex \ac{SDP} that can be solved by convex solver tools such as CVX~\cite{grant2014cvx}. After solving $\mathcal{P}_3$, we obtain $\bm{W}^n$ as the optimal solution for the current iteration $n$. If $\bm{W}^n$ is a rank-one matrix, we can use Cholesky decomposition to derive $\bm{x}^n$ as the optimal modulation vector solution for $\mathcal{P}_1$~\cite{saeed2023ChannelComp}. Otherwise, a sub-optimal solution for $\mathcal{P}_1$ can be obtained through a rank-one approximation of $\bm{W}^n$.

\subsubsection{Optimize for the {\color{black}Coding} Vector}

Problem $\mathcal{P}_2$ in~\eqref{eq:MBO} is a \ac{MIQCP} problem with a quadratic objective and quadratic constraints. To manage its non-convexity and NP-hardness, we employ the relaxation technique proposed in~\cite{zhao2017global} to transform the problem into a relaxed \ac{LP} problem. Specifically, this relaxation simplifies the problem, making it more tractable and easier to solve using the MATLAB CVX toolbox~\cite{grant2014cvx}. The convexity of the resulting \ac{LP} ensures that alternating minimization converges to a stationary point. Moreover, this relaxation provides a lower bound on the optimal value of Problem $\mathcal{P}_2$, as stated in the following proposition.

\begin{prop}\label{pro:linear relaxation}
For Problem $\mathcal{P}_{2}$, there exists a linear relaxation programming
problem $\mathcal{P}_{4}$ that provides a valid lower bound for the optimal value of $\mathcal{P}_2$:
\begin{align}  \nonumber
\mathcal{P}_2  \approx  \mathcal{P}_{4} &:=  \min_{\bm{c}_{\ell}}  \quad \sum_{\ell=1}^L \mathds{1}^{\top}\bm{c}_{\ell}, \nonumber \\  \nonumber
{\rm s.t.} & \sum_{\ell=1}^L \sum_{m=1}^{r_{ij}} 
2l_{m}^{i,j}\bm{p}_{m}^{i,j\top}\bm{c}_{\ell} \leq -\Delta f_{i,j} + \sum_{\ell=1}^L \sum_{m=1}^{r_{ij}}(l_{m}^{i,j})^2, \\  \nonumber
& \sum_{\ell=1}^L\sum_{m=1}^{r_{ij}} 
2u_{m}^{i,j}\bm{p}_{m}^{i,j\top}\bm{c}_{\ell} \leq -\Delta f_{i,j} + \sum_{\ell=1}^L \sum_{m=1}^{r_{ij}}(u_{m}^{i,j})^2, \\ \label{eq:relaxation_programming_high}
& \bm{c}_\ell \in [0,1]^N, \quad \forall \ell \in [L].
\end{align}
Here, $\bm{p}_{m}^{i,j}$ represent the decomposed vectors of the matrix $\bm{P}^{i,j}:=(\bm{a}_i-\bm{a}_j) (\bm{a}_j-\bm{a}_i)^{\top} \odot \bm{W}^n$, where rank$(\bm{P}^{i,j})=r_{ij}$. Additionally, let $\bm{l}^{i,j}:=[l_1^{i,j}, \ldots, l_{r_{ij}}^{i,j}]$ and $\bm{u}^{i,j}:=[u_1^{i,j}, \ldots, u_{r_{ij}}^{i,j}]$ denote the vectors of lower and upper bounds for the bilinear terms in the constraints. Therefore, the optimality gap of each constraint between Problem $\mathcal{P}_2$ and $\mathcal{P}_4$ can be upper bounded by $\Delta h^{i,j}(\bm{C}) = L\|\bm{u}^{i,j}-\bm{l}^{i,j}\|_2^2$. 
\end{prop} 

\begin{proof}
See Appendix~\ref{app:linear relaxation}.
\end{proof}

Proposition~\ref{pro:linear relaxation} establishes that the optimal value of Problem $\mathcal{P}_{4}$ provides a lower bound for the global optimal value of Problem $\mathcal{P}_{2}$. Since both $\mathcal{P}_3$ and $\mathcal{P}_4$ are convex optimization problems, iteratively minimizing over $\bm{W}$ and $\bm{C}$ leads to convergence at a stationary point $(\bm{W}^n, \bm{C}^n)$, serving as the optimal solutions for $\mathcal{P}_3$ and $\mathcal{P}_4$.

\subsubsection{Project Back to Feasible Sets}

After reaching the stationary point $(\bm{W}^n, \bm{C}^n)$, if all $\bm{c}_{\ell}^n$s are binary vectors and $\bm{W}^n$ is a rank-one matrix, then $(\bm{W}^n, \bm{C}^n)$ is considered as the feasible solution $(\hat{\bm{W}}, \hat{\bm{C}})$ for $\mathcal{P}_1$ and $\mathcal{P}_2$.  Otherwise, we need to project it back to the feasible set of the primal optimization problems $\mathcal{P}_1$ and $\mathcal{P}_2$. On the one hand, we use Cholesky decomposition for $\bm{W}^n$ to obtain an approximate solution $\hat{\bm{x}}$. More precisely, $\hat{\bm{x}}$ is obtained as follows 
\begin{align} \label{eq:Cholesky decomposition}
    \hat{\bm{x}} = \sqrt{\lambda_1}\bm{u}_1^n,
\end{align}
where $\lambda_1 \geq 0$ is the largest eigenvalue of $\bm{W}^n$, and   $\bm{u}_1^n\in \mathbb{C}^{N\times 1}$ is the eigenvector corresponding to $\lambda_1$, which is a unitary vector\footnote{Recall that the constraint in \eqref{eq:lift-rank-relaxed} ensures that the matrix $\bm{W}^n$  is positive semi-definite, which means all its eigenvalues are non-negative, i.e., $\lambda_i \geq 0$ for $i\in [N]$.}. On the other hand, we use the branch and bound algorithm~\cite{linderoth2005simplicial} to obtain binary variables $\hat{\bm{c}}_{\ell}$s that satisfy the constraints of Problem $\mathcal{P}_{2}$. This approach provides a tight lower bound for the global optimum of Problem $\mathcal{P}_{2}$. Practically, we can obtain $\hat{\bm{c}}_{\ell}$s by using the Gurobi toolbox~\cite{pedroso2011optimization}.

\begin{rem}\label{pro:MIQCPconvergence}
Note that the branch and bound algorithm either terminates within finite iterations to a global optimal solution for Problem $\mathcal{P}_{2}$, or determines that Problem $\mathcal{P}_2$ is infeasible. The approach to finding the optimal solution is reported in Appendix~\ref{app:MIQCPconvergence}. 
\end{rem}

The complete procedure of the alternating minimization approach is summarized in Algorithm~\ref{Alg:HIT}. Codes for implementing this algorithm are provided on Github\footnote{ https://github.com/xiaojingyan-elsa/ReMAC.git.}. In Subsection~\ref{sec:Theory}, we provide the theoretical guarantee regarding Algorithm~\ref{Alg:HIT}.


\begin{algorithm}[!t]\label{Alg:HIT}
    \caption{ReMAC Algorithm}
    
    \textbf{Initialize} $\bm{W}^{0}:= \bm{x}^{0} \bm{x}^{0 \mathsf{H}}$, $\bm{C}^{0}$, the maximum number of iteration $T$, and set the current iteration counter as $n = 1$. 
    
    \textbf{Repeat}

        \Indp
        Update $\bm{W}$: Obtain $\bm{W}^{n}$ by solving $\mathcal{P}_{3}$. \\
    
        Update $\bm{C}$: Obtain $\bm{C}^{n}$ by solving $\mathcal{P}_{4}$.
        
        Increment the iteration counter: $n \leftarrow n + 1$.
        
        \Indm
        
        \textbf{Until} $||\bm{C}^{n}-\bm{C}^{n-1}||_{\rm F} \leq \Delta$ or $n=T$.

        Use Cholesky decomposition to obtain the solution $\hat{\bm{x}}$ for $\mathcal{P}_1$.

        Use branch and bound to obtain the solution $\hat{\bm{C}}$ for $\mathcal{P}_2$.

    \textbf{Output} approximate solution $(\hat{\bm{x}},\hat{\bm{C}})$ for $\mathcal{P}_0$.
 
\end{algorithm}


\subsection{Theoretical Guarantees}\label{sec:Theory}

In this subsection, we analyze the convergence of Algorithm~\ref{Alg:HIT} to assess its speed and the proximity of the obtained solution to the optimal one. 
Based on Subsection~\ref{sec:MethodDesign}, we know that the relaxed Problems $\mathcal{P}_3$ and $\mathcal{P}_4$ provide lower bounds for the optimal values of Problems $\mathcal{P}_1$ and $\mathcal{P}_2$. This allows us to analyze the gap between the solutions of the relaxed and the original problems. Additionally, since both $\mathcal{P}_3$ and $\mathcal{P}_4$ are convex, alternating minimization between them converges to a first-order stationary point~\cite{li2019alternating}.
Therefore, to further illustrate the efficiency of our approach, we present the convergence rate of Algorithm~\ref{Alg:HIT} in the following proposition.

\begin{prop}\label{theorem:convergence rate}
Let $\{\bm{x}^{n},\bm{c}_{\ell}^{n}\}_{n\geq 0}$ be the sequence generated by Algorithm~\ref{Alg:HIT}. Then, the gap between the objective value of Problem $\mathcal{P}_2$ using this sequence and the optimal value of Problem $\mathcal{P}_2$ is bounded as  
\begin{align}
\sum_{\ell=1}^L\|\bm{c}_{\ell}^{n}\|_1 - \sum_{\ell=1}^L\|\bm{c}_{\ell}^{*}\|_1 \leq \frac{2R(\bm{x}^{0},\bm{c}_{\ell}^{0})}{n-1}, \quad n\geq 2,
\label{eq:convergence_rate}
\end{align}
where  {\color{black}
\begin{align*}  
R(\bm{x}^0,\bm{C}^0) :=& \max_{(\bm{x},\bm{C})\in \mathcal{S}}\Big\{ R_{1}(\bm{x},\bm{x}^*) + R_{2}(\bm{C},\bm{C}^*) \Big\},  \\
& {\rm s.t.} \sum_{\ell=1}^L\|\bm{c}_{\ell}\|_1 \leq \sum_{\ell=1}^L\|\bm{c}_{\ell}^{0}\|_1,
\end{align*}
with  
\begin{align} \nonumber  
    R_{1}(\bm{x},\bm{x}^*)& := (2\|\bm{x}-\bm{x}^*\|_{\rm 2}+ 4\sqrt{\tilde{M}})^2, \\ \nonumber  
    R_{2}(\bm{C},\bm{C}^*) & := \sum_{\ell=1}^L(\|\bm{c}_{\ell}-\bm{c}_{\ell}^*\|_{\rm 2} + 2\|\bm{c}_{\ell}^*\|_{\rm 2})^2.  
\end{align}
}Here, $\mathcal{S}$ denotes the feasible set of Problems $\mathcal{P}_3$ and $\mathcal{P}_4$ involving all possible values of the variables $\bm{x}$ and $\bm{c}_{\ell}$s for $\ell \in [L]$. The variables $\bm{x}^*$ and $\bm{c}_{\ell}^*$s represent the optimal solutions for Problems $\mathcal{P}_1$ and $\mathcal{P}_2$, respectively. Additionally, $\tilde{M}$ indicates the number of constraints in Problem $\mathcal{P}_0$.  
\end{prop}


\begin{proof}
See Appendix~\ref{app:convergence rate}.
\end{proof}

According to Proposition~\ref{theorem:convergence rate}, the gap between the objective value \( \sum_{\ell=1}^L\|\bm{c}_{\ell}^{n}\|_1 \) at each iteration \( n \) and the optimal value \( \sum_{\ell=1}^L\|\bm{c}_{\ell}^{*}\|_1 \) narrows as \( n \) increases, indicating that the performance of Algorithm~\ref{Alg:HIT} improves over successive iterations. Meanwhille, the convergence rate of Algorithm~\ref{Alg:HIT} is influenced by the value of \( R(\bm{x}^{0}, \bm{c}_{\ell}^{0}) \). {\color{black}Precisely, \( R_{1}(\bm{x},\bm{x}^*) \) bounds the difference between the obtained solution \( \bm{x} \) and the optimal solution \( \bm{x}^* \) for subproblem \( \mathcal{P}_1 \), while \( R_{2}(\bm{C},\bm{C}^*) \) measures the gap between the obtained solution \( \bm{C} \) and the optimal solution \( \bm{C}^* \) for subproblem \( \mathcal{P}_2 \). Additionally, \( R(\bm{x}^{0}, \bm{c}_{\ell}^{0}) \) is related to the number of constraints \( \tilde{M} \) in Problem \( \mathcal{P}_{0} \).} Since \( \tilde{M} \) depends on the specific computation function \( f \) and increases with the number of input values \( Q \) or nodes \( K \) in the network, Algorithm~\ref{Alg:HIT} tends to converge more slowly when handling a larger number of nodes or high-order modulation schemes.


{\color{black} \subsection{Complexity Analysis} \label{subsec:complexity}}

{\color{black}In this subsection, we analyze the computational complexity associated with Algorithm~\ref{Alg:HIT} when applied to solving Problem $\mathcal{P}_0$ for computing desired function $f$. To analyze the computational complexity for Problem $\mathcal{P}_0$, there are four main factors that play a role in scaling the complexity. These factors include the number of nodes \( K \), the quantization level \( Q \), the number of time slots \( L \), and the number of constraints \( \tilde{M} \).  Since the algorithm alternates between optimizing the modulation vector \( \bm{x} \) and the repetition coding matrix \( \bm{C} \), we need to consider the complexities of the two subproblems, \(\mathcal{P}_3\) and \(\mathcal{P}_4\), separately.}

{\color{black}In each iteration, the first step requires solving the  SDP in \(\mathcal{P}_3\) to obatin the modulation vector \( \bm{x} \). To solve  SDP with the CVX package, the toolbox employs an interior-point algorithm~\cite{ye2011interior}, whose complexity scales polynomially with the size of the problem and the number of constraints. More preciscly, the complexity of the  SDP scales as \( \mathcal{O}(\tilde{M}^4N^{0.5}) \), and by substituting $N$ with $QK$, it becomes \( \mathcal{O}(\tilde{M}^4Q^{0.5}K^{0.5})\).

In the second step, the algorithm optimizes the repetition coding matrix \( \bm{C}\) by solving the relaxed  LP in \(\mathcal{P}_4\). The computational complexity of the  LP is \( \mathcal{O}(\tilde{M}^3Q^{0.5}K^{0.5}L^{0.5}) \)~\cite{antoniou2007practical}. Hence, considering \( N_{iter} \) as the number of iterations required for the algorithm to converge, the overall complexity of the alternating minimization procedure becomes \( \mathcal{O}(N_{iter}\tilde{M}^3Q^{0.5}K^{0.5}(\tilde{M} + L^{0.5})) \).}

{\color{black}After convergence, we must project the lifted modulation matrix and the repetition encoding matrix back onto the feasible set by Cholesky decomposition and the branch and bound method, respectively. Accordingly, the Cholesky decomposition requires a complexity of \( \mathcal{O}(Q^3K^3) \)~\cite{trefethen2022numerical}, and the branch and bound with a worst-case complexity of \( \mathcal{O}(2^{KQL}) \)~\cite{zhang1996branch}. However, we highlight that efficient pruning of non-promising paths for the branch and bound method reduces the search space and computational cost. Indeed, modern solvers such as Gurobi further enhance computational efficiency with advanced heuristics and parallel processing, ensuring the method remains practical for larger values of \( K \), \( Q \) and~\( L \).}

{\color{black}\begin{rem}\label{pro:scalability}
 Note that \ac{ReMAC} is designed as an offline framework, requiring the optimization problems to be solved only once during the system initial setup. Therefore, the computational complexity of ReMAC is feasible to manage for-world wireless systems. Indeed, the resultant modulation vector and repetition coding matrix can be stored and reused during real-time operation. Hence, this reusability eliminates computational burdens and ensures suitability for resource-constrained applications.
\end{rem}}

In the next section, we empirically check the performance of the proposed \ac{ReMAC}.

\section{Numerical Experiments}\label{sec:Num}

In this section, we evaluate the performance of \ac{ReMAC} for different function computations and make a comparison with other methods. We provide the numerical results of \ac{ReMAC} for computing the sum $f=\sum_{k=1}^Kx_k$, product $f=\prod_{k=1}^Kx_k$ and max $f=\max_{k}x_k$ functions over the \ac{MAC} under various \ac{SNR} levels and coding time slots in terms of the \ac{NMSE} metric. Moreover, we compare \ac{ReMAC} with the following {\color{black} three} different methods:
\begin{itemize}
    \item ChannelComp~\cite{razavikia2023computing}: we use the digital modulation vectors designed by ChannelComp. For comparison, the same modulated symbols are transmitted multiple times across $L$ consecutive time slots. The \ac{CP} then averages the recovered output values over each time slot to derive the final output. For simplicity in notation, we use \ac{ReMAC} $L=1$ to represent ChannelComp.
    \item {\color{black} Bit-slicing~\cite{liu2024digital}: we use the bit slicing framework by spliting discrete input values into $L$ bit segments and transmitting each segment as a modulated symbol over $L$ consecutive time slots. The \ac{CP} subsequently assembles these symbols to reconstruct the output value.} 
    \item {\color{black}Digital \ac{AirComp}~\cite{zhao2022broadband}: we use the traditional \ac{AirComp} framework, where discrete input values are modulated using \ac{PAM}. At the \ac{CP}, the received signals are averaged over multiple transmissions to estimate the output of the target function.}
\end{itemize}  

\begin{figure*}[!t]
\centering
\subfigure[$f=\sum_{k=1}^4x_k$]{\label{fig:four_node_sum_SNR}
\begin{tikzpicture}
    \begin{axis}[
        xlabel = {SNR(dB)},
        ylabel = {NMSE},
        label style={font=\footnotesize},
        width=0.32\textwidth,
        height=5cm,
        xmin=15, xmax=40,
        ymin=0.05, ymax=0.15,
        legend style={nodes={scale=0.65, transform shape}, at={(0.3,0.85)}},
        ticklabel style = {font=\footnotesize},
        legend pos=south west,  
        ymajorgrids=true,
        xmajorgrids=true,
        grid style=dashed,
        grid=both,
        ymode = log,
        grid style={line width=.1pt, draw=gray!10},
        major grid style={line width=.2pt,draw=gray!30},
    ]
    \addplot[
             thin,
        color=chestnut,
        mark=*,
        line width=0.9pt,
        mark size=1.5pt,
        ]
    table[x=SNR,y=L1]
    {Data/Sim_L_8_bits.dat};
    \addplot[ 
            color=airforceblue,
            mark=square*,
            line width=0.9pt,
            mark size=1.5pt,
            ]
    table[x=SNR,y=L2]
    {Data/Sim_L_8_bits.dat};
    \addplot[ 
        color=cssgreen,
        mark=triangle*,
        line width=0.9pt,
        mark size=1.5pt,
        ]
    table[x=SNR,y=L3]
    {Data/Sim_L_8_bits.dat};
    \addplot[ 
        color=cadmiumorange,
        mark=diamond*,
        line width=0.9pt,
        mark size=1.5pt,
        ]
    table[x=SNR,y=L4]
    {Data/Sim_L_8_bits.dat};
    \legend{$L=1$, $L=2$, $L=3$, $L=4$};
\end{axis}
\end{tikzpicture}}
\subfigure[$f = \prod_{k=1}^4x_k$]{   \label{fig:four_node_product_SNR}
\begin{tikzpicture}
    \begin{axis}[
        xlabel = {SNR(dB)},
        ylabel = {NMSE},
        label style={font=\footnotesize},
        width=0.32\textwidth,
        height=5cm,
        xmin=15, xmax=40,
        ymin=0.35, ymax=2.1,
        legend style={nodes={scale=0.65, transform shape}, at={(0.008,0.0006)} },
        ticklabel style = {font=\footnotesize},
        legend pos=south west,
        ymajorgrids=true,
        xmajorgrids=true,
        grid style=dashed,
        grid=both,
        ymode = log,
        grid style={line width=.1pt, draw=gray!10},
        major grid style={line width=.2pt,draw=gray!30},
    ]
    \addplot[
             thin,
        color=chestnut,
        mark=*,
        line width=0.9pt,
        mark size=1.5pt,
        ]
    table[x=SNR,y=L1p]
    {Data/Sim_L_8_bits.dat};
    \addplot[ 
            color=airforceblue,
            mark=square*,
            line width=0.9pt,
            mark size=1.5pt,
            ]
    table[x=SNR,y=L2p]
    {Data/Sim_L_8_bits.dat};
    \addplot[ 
        color=cssgreen,
        mark=triangle*,
        line width=0.9pt,
        mark size=1.5pt,
        ]
    table[x=SNR,y=L3p]
    {Data/Sim_L_8_bits.dat};
    \addplot[ 
        color=cadmiumorange,
        mark=diamond*,
        line width=0.9pt,
        mark size=1.5pt,
        ]
    table[x=SNR,y=L4p]
    {Data/Sim_L_8_bits.dat};
    \legend{$L=1$, $L=2$,$L=3$, $L=4$};
\end{axis}
\end{tikzpicture}
}\subfigure[$f=\max_{k}x_k$]{\label{fig:four_node_max_SNR}
\begin{tikzpicture}
    \begin{axis}[
        xlabel = {SNR(dB)},
        ylabel = {NMSE},
        label style={font=\footnotesize},
        width=0.32\textwidth,
        height=5cm,
        xmin=15, xmax=40,
        ymin=0.01, ymax=0.1,
        legend style={nodes={scale=0.65, transform shape}, at={(0.3,0.85)}},
        ticklabel style = {font=\footnotesize},
        legend pos=south west,  
        ymajorgrids=true,
        xmajorgrids=true,
        grid style=dashed,
        grid=both,
        ymode = log,
        grid style={line width=.1pt, draw=gray!10},
        major grid style={line width=.2pt,draw=gray!30},
    ]
    \addplot[
             thin,
        color=chestnut,
        mark=*,
        line width=0.9pt,
        mark size=1.5pt,
        ]
    table[x=SNR,y=L1m]
    {Data/Sim_L_8_bits.dat};
    \addplot[ 
            color=airforceblue,
            mark=square*,
            line width=0.9pt,
            mark size=1.5pt,
            ]
    table[x=SNR,y=L2m]
    {Data/Sim_L_8_bits.dat};
    \addplot[ 
        color=cssgreen,
        mark=triangle*,
        line width=0.9pt,
        mark size=1.5pt,
        ]
    table[x=SNR,y=L3m]
    {Data/Sim_L_8_bits.dat};
    \addplot[ 
        color=cadmiumorange,
        mark=diamond*,
        line width=0.9pt,
        mark size=1.5pt,
        ]
    table[x=SNR,y=L4m]
    {Data/Sim_L_8_bits.dat};
    \legend{$L=1$, $L=2$, $L=3$, $L=4$};
\end{axis}
\end{tikzpicture}}
  \caption{Performance of ReMAC under different SNRs in terms of NMSE averaged over $N_s=100$. The input values are $x_k \in \{1,2,\ldots,256\}$ and the desired functions are $f = \sum_{k=1}^Kx_k$, $f = \prod_{k=1}^Kx_k$ and $f=\max_{k}x_k$ with $K=4$ nodes in the network.}
  \label{fig:four_node_product_SN}
\end{figure*}
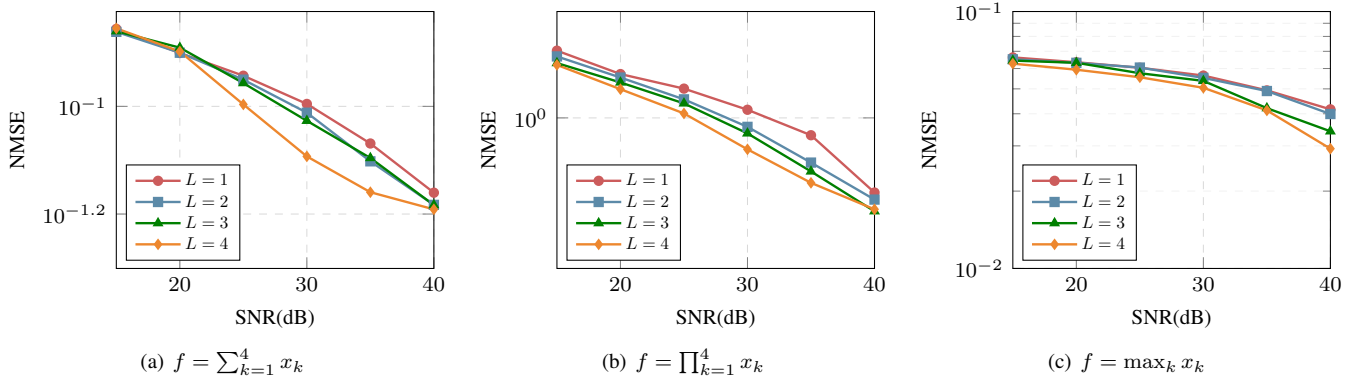

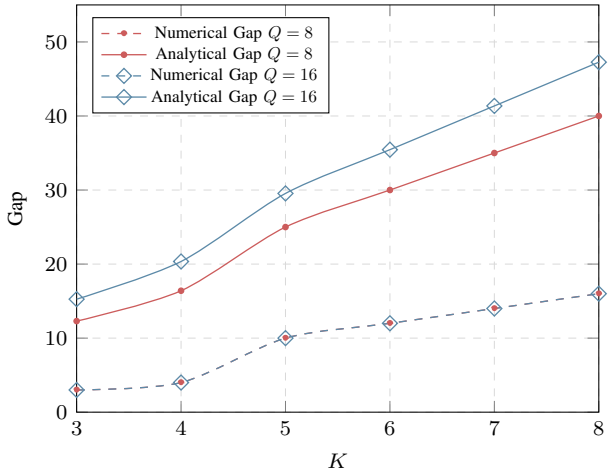
\begin{figure}[!t]
\centering
\begin{tikzpicture}[
    domain=0:4,
    spy using outlines={circle, magnification=3, size=1.5cm, connect spies, every spy on node/.append style={thick}}, 
    ]
    \begin{axis}[
        xlabel = {$K$},
        ylabel = {Gap},
        label style={font=\footnotesize},
        width=0.47\textwidth,
        height=7cm,
        xmin=3, xmax=8,
        ymin=0, ymax=55,
        legend style={nodes={scale=0.65, transform shape}, at={(0.3,0.85)}},
        ticklabel style = {font=\footnotesize},
        legend pos=north west,
        ymajorgrids=true,
        xmajorgrids=true,
        grid style=dashed,
        grid=both,
        grid style={line width=.1pt, draw=gray!10},
        major grid style={line width=.2pt,draw=gray!30},
    ]
    \addplot[smooth,
             thin,
             dashed,
        color=chestnut,
        mark=*,
        line width=0.5pt,
        mark size=1pt,
        ]
    table[x=K,y=gap_3]
    {Data/Sim_objective_prod.dat};
    \addplot[smooth,
              thin,
            color=chestnut,
            mark=*,
            line width=0.5pt,
            mark size=1pt,
            ]
    table[x=K,y=upper_gap_3]
    {Data/Sim_objective_prod.dat};
    \addplot[smooth,
             thin, dashed,
        color=airforceblue,
        mark=square,
        mark options = {rotate = 45,solid},
        line width=0.5pt,
        mark size=2pt,
        ]
    table[x=K,y=gap_4]
    {Data/Sim_objective_prod.dat};
    \addplot[ smooth,
             thin,
        color=airforceblue,
        mark=square,
        mark options = {rotate = 45,solid},
        line width=0.5pt,
        mark size=2pt,
        ]
    table[x=K,y=upper_gap_4]
    {Data/Sim_objective_prod.dat};
\legend{Numerical Gap $Q=8$, Analytical Gap $Q=8$, Numerical Gap $Q=16$, Analytical Gap $Q=16$};
\end{axis}
\end{tikzpicture}
\caption{Comparison of the optimality gap between numerical results and proximity analysis from Proposition 2. The numerical gap is expressed as $\sum_{\ell=1}^L\|\bm{c}_{\ell}^{n}\|_1 - \sum_{\ell=1}^L\|\bm{c}_{\ell}^{*}\|_1$, while the analytical gap is give by $2R_{2}(\bm{C},\bm{C}^*)/(n-1)$. The network size varies from \(K=3\) to \(K=8\) with quantization levels \(Q \in \{8, 16\}\). The number of time slots is \(L=2\), and the computed function is \(f = \prod_{k=1}^K x_k\). }
\label{fig:optimality_gaps}
\end{figure}

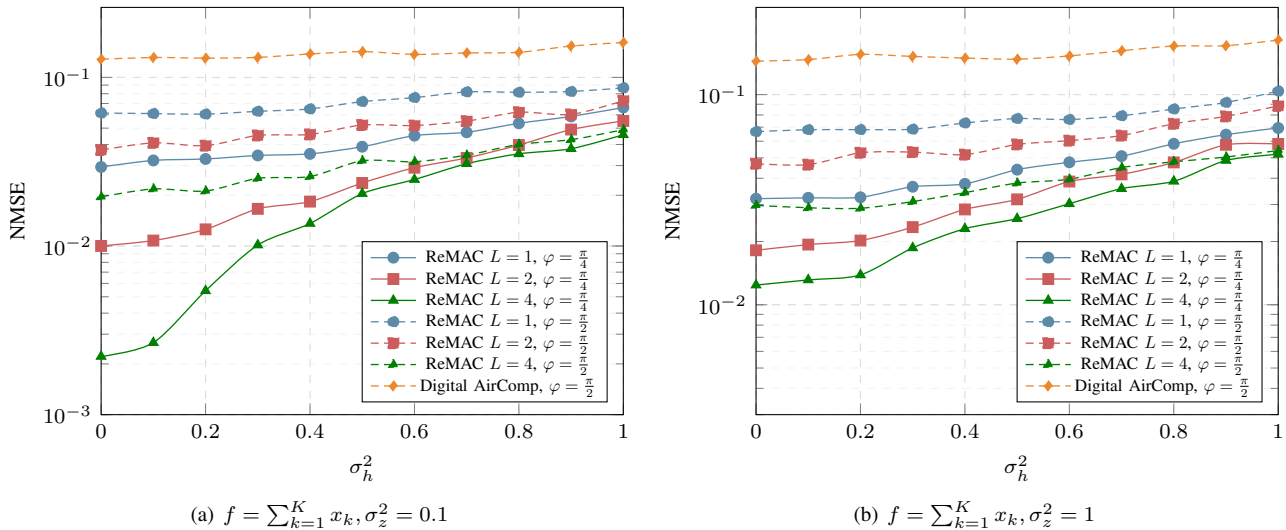
\begin{figure*}[!t]
\label{fig:fading variance sum}
\centering
\subfigure[$f = \sum_{k=1}^Kx_k,\sigma_z^2=0.1$]{\label{fig:four_node_sum_NMSE}
\begin{tikzpicture}[
    domain=0:4,
    spy using outlines={circle, magnification=3, size=2.4cm, connect spies, every spy on node/.append style={thick}}, 
    ]
    \begin{axis}[
        xlabel = {$\sigma_h^2$},
        ylabel = {NMSE},
        label style={font=\footnotesize},
        width=0.47\textwidth,
        height=7cm,
        xmin=0, xmax=1,
        ymin=0.001, ymax=0.26,
        legend style={nodes={scale=0.65, transform shape}, at={(0.3,0.85)}},
        ticklabel style = {font=\footnotesize},
        legend pos=south east,
        ymajorgrids=true,
        xmajorgrids=true,
        grid style=dashed,
        grid=both,
        ymode = log,
        grid style={line width=.1pt, draw=gray!10},
        major grid style={line width=.2pt,draw=gray!30},
    ]
    \addplot[smooth,
             thin,
        color=airforceblue,
        mark=*,
        line width=0.5pt,
        mark size=2pt,
        ]
    table[x=sigmah,y=L1]
    {Data/Sim_fading_sum.dat};
    \addplot[smooth,
              thin,
            color=chestnut,
            mark=square*,
            line width=0.5pt,
            mark size=2pt,
            ]
    table[x=sigmah,y=L2]
    {Data/Sim_fading_sum.dat};
    \addplot[smooth,
             thin,
        color=cssgreen,
        mark=triangle*,
        line width=0.5pt,
        mark size=2pt,
        ]
    table[x=sigmah,y=L4]
    {Data/Sim_fading_sum.dat};
    \addplot[smooth,
             thin,
             densely dashed,
        color=airforceblue,
        mark=*,
        line width=0.5pt,
        mark size=2pt,
        ]
    table[x=sigmah,y=L1p]
    {Data/Sim_fading_sum.dat};
    \addplot[ smooth,
              thin,
              densely dashed,
            color=chestnut,
            mark=square*,
            line width=0.5pt,
            mark size=2pt,
            ]
    table[x=sigmah,y=L2p]
    {Data/Sim_fading_sum.dat};
    \addplot[ smooth,
             thin,
             densely dashed,
        color=cssgreen,
        mark=triangle*,
        line width=0.5pt,
        mark size=2pt,
        ]
    table[x=sigmah,y=L4p]
    {Data/Sim_fading_sum.dat};
    \addplot[ smooth,
             thin,
             densely dashed,
        color=cadmiumorange,
        mark=diamond*,
        line width=0.5pt,
        mark size=2pt,
        ]
    table[x=sigmah,y=aircompp]
    {Data/Sim_fading_sum.dat};
\legend{ReMAC $L=1\text{, }\varphi=\frac{\pi}{4}$, ReMAC $L=2\text{, }\varphi=\frac{\pi}{4}$, ReMAC $L=4\text{, }\varphi=\frac{\pi}{4}$, ReMAC $L=1\text{, }\varphi=\frac{\pi}{2}$, ReMAC $L=2\text{, }\varphi=\frac{\pi}{2}$, ReMAC $L=4\text{, }\varphi=\frac{\pi}{2}$, {\color{black}Digital }AirComp$\text{, }\varphi=\frac{\pi}{2}$};
\end{axis}
\end{tikzpicture}
  }\subfigure[$f=\sum_{k=1}^Kx_k,\sigma_z^2=1$]{\label{fig:four_node_prod_NMSE}
\centering
\begin{tikzpicture}[
    domain=0:4,
    spy using outlines={circle, magnification=3, size=1.8cm, connect spies, every spy on node/.append style={thick}}, 
    ]
    \begin{axis}[
        xlabel = {$\sigma_h^2$},
        ylabel = {NMSE},
        label style={font=\footnotesize},
        width=0.47\textwidth,
        height=7cm,
        xmin=0, xmax=1,
        ymin=0.003, ymax=0.26,
        legend style={nodes={scale=0.65, transform shape}, at={(0.3,0.85)}},
        ticklabel style = {font=\footnotesize},
        legend pos=south east,
        ymajorgrids=true,
        xmajorgrids=true,
        grid style=dashed,
        grid=both,
        ymode = log,
        grid style={line width=.1pt, draw=gray!10},
        major grid style={line width=.2pt,draw=gray!30},
    ]
    \addplot[smooth,
             thin,
        color=airforceblue,
        mark=*,
        line width=0.5pt,
        mark size=2pt,
        ]
    table[x=sigmah,y=L1_h]
    {Data/Sim_fading_sum.dat};
    \addplot[smooth,
              thin,
            color=chestnut,
            mark=square*,
            line width=0.5pt,
            mark size=2pt,
            ]
    table[x=sigmah,y=L2_h]
    {Data/Sim_fading_sum.dat};
    \addplot[smooth,
             thin,
        color=cssgreen,
        mark=triangle*,
        line width=0.5pt,
        mark size=2pt,
        ]
    table[x=sigmah,y=L4_h]
    {Data/Sim_fading_sum.dat};
    \addplot[smooth,
             thin,
             densely dashed,
        color=airforceblue,
        mark=*,
        line width=0.5pt,
        mark size=2pt,
        ]
    table[x=sigmah,y=L1p_h]
    {Data/Sim_fading_sum.dat};
    \addplot[ smooth,
              thin,
              densely dashed,
            color=chestnut,
            mark=square*,
            line width=0.5pt,
            mark size=2pt,
            ]
    table[x=sigmah,y=L2p_h]
    {Data/Sim_fading_sum.dat};
    \addplot[ smooth,
             thin,
             densely dashed,
        color=cssgreen,
        mark=triangle*,
        line width=0.5pt,
        mark size=2pt,
        ]
    table[x=sigmah,y=L4p_h]
    {Data/Sim_fading_sum.dat};
    \addplot[ smooth,
             thin,
             densely dashed,
        color=cadmiumorange,
        mark=diamond*,
        line width=0.5pt,
        mark size=2pt,
        ]
    table[x=sigmah,y=aircompp_h]
    {Data/Sim_fading_sum.dat};
\legend{ReMAC $L=1\text{, }\varphi=\frac{\pi}{4}$, ReMAC $L=2\text{, }\varphi=\frac{\pi}{4}$, ReMAC $L=4\text{, }\varphi=\frac{\pi}{4}$, ReMAC $L=1\text{, }\varphi=\frac{\pi}{2}$, ReMAC $L=2\text{, }\varphi=\frac{\pi}{2}$, ReMAC $L=4\text{, }\varphi=\frac{\pi}{2}$, {\color{black}Digital }AirComp$\text{, }\varphi=\frac{\pi}{2}$};
\end{axis}
\end{tikzpicture}
}
  \caption{Performance comparison between \ac{ReMAC}, ChannelComp and {\color{black}Digital} \ac{AirComp} in the presence of fading channels among $K=8$ nodes. The \ac{NMSE} is depicted vesus the variance of channel coefficients $\sigma_h^2$ under different transmission slots, i.e., $L \in \{1,2,4\}$, and phase shifts, i.e., {\color{black}$\varphi \in \{\pi/4,\pi/2\}$}. Input values are $x_k \in \{1,2,3,4\}$ and the desired function is $f = \sum_{k=1}^Kx_k$, respectively. The noise follows a normal distribution with low variance $\sigma_z^2=0.1$ and high variance {\color{black}$\sigma_z^2=1$}. 
  }\label{fig:four_node_sum_NMS}
\end{figure*}

The metric \ac{NMSE} is defined as the sum of the squared differences between desired function values $f^{(i)}$ and their estimated counterparts $\hat{f}_j^{(i)}$ normalized by the number of Monte Carlo trials $N_s$ times the square of the absolute value of $f^{(i)}$:
\begin{align}
\text{\ac{NMSE}}:= \frac{1}{N_s}\frac{\sum\nolimits_{j=1}^{N_s}|f^{(i)}-\hat{f}_j^{(i)}|^2}{|f^{(i)}|^2}.
\end{align}
We also define \ac{SNR}$:=10\log(\|\bm{x}\|_2^2/\sigma_z^2)$, where $\sigma_z^2$ is the variance of the noise $\vec{\tilde{z}}_{\ell}$ in \eqref{eq:nofading}. Note that the variance of $\vec{\tilde{z}}_{\ell}$ equals to the minimum MSE of the corresponding power
control scheme in~\cite{cao2020optimal}.

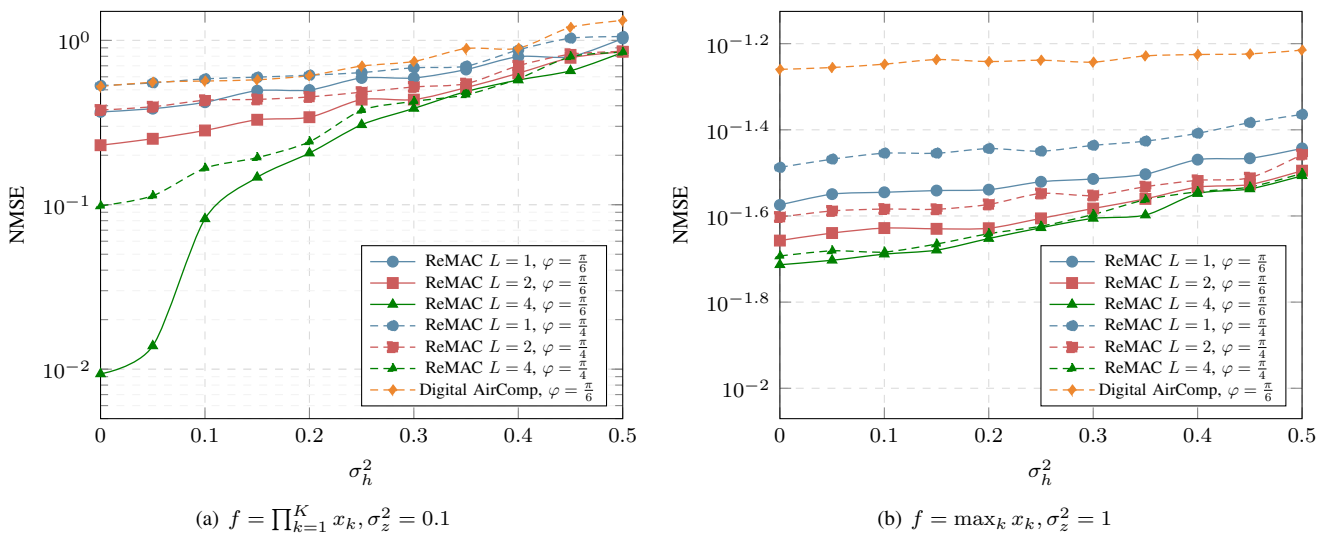
\begin{figure*}[!t]
\centering
\subfigure[$f = \prod_{k=1}^Kx_k,\sigma_z^2=0.1$]{\label{fig:fading_prod_NMSE}
\begin{tikzpicture}[
    domain=0:4,
    spy using outlines={circle, magnification=3, size=2cm, connect spies, every spy on node/.append style={thick}}, 
    ]
    \begin{axis}[
        xlabel = {$\sigma_h^2$},
        ylabel = {NMSE},
        label style={font=\footnotesize},
        width=0.47\textwidth,
        height=7cm,
        xmin=0, xmax=0.5,
        ymin=0.005, ymax=1.5,
        legend style={nodes={scale=0.65, transform shape}, at={(0.3,0.85)}},
        ticklabel style = {font=\footnotesize},
        legend pos=south east,
        ymajorgrids=true,
        xmajorgrids=true,
        grid style=dashed,
        grid=both,
        ymode = log,
        grid style={line width=.1pt, draw=gray!10},
        major grid style={line width=.2pt,draw=gray!30},
    ]
    \addplot[smooth,
             thin,
        color=airforceblue,
        mark=*,
        line width=0.5pt,
        mark size=2pt,
        ]
    table[x=sigmah,y=L1]
    {Data/Sim_fading_prod.dat};
    \addplot[smooth,
              thin,
            color=chestnut,
            mark=square*,
            line width=0.5pt,
            mark size=2pt,
            ]
    table[x=sigmah,y=L2]
    {Data/Sim_fading_prod.dat};
    \addplot[smooth,
             thin,
        color=cssgreen,
        mark=triangle*,
        line width=0.5pt,
        mark size=2pt,
        ]
    table[x=sigmah,y=L4]
    {Data/Sim_fading_prod.dat};
    \addplot[smooth,
             thin,
             densely dashed,
        color=airforceblue,
        mark=*,
        line width=0.5pt,
        mark size=2pt,
        ]
    table[x=sigmah,y=L1p]
    {Data/Sim_fading_prod.dat};
    \addplot[ smooth,
              thin,
            densely dashed,
            color=chestnut,
            mark=square*,
            line width=0.5pt,
            mark size=2pt,
            ]
    table[x=sigmah,y=L2p]
    {Data/Sim_fading_prod.dat};
    \addplot[ smooth,
             thin,
        densely dashed,
        color=cssgreen,
        mark=triangle*,
        line width=0.5pt,
        mark size=2pt,
        ]
    table[x=sigmah,y=L4p]
    {Data/Sim_fading_prod.dat};
    \addplot[ smooth,
             thin,
        densely dashed,
        color=cadmiumorange,
        mark=diamond*,
        line width=0.5pt,
        mark size=2pt,
        ]
    table[x=sigmah,y=aircomp]
    {Data/Sim_fading_prod.dat};
\legend{ReMAC $L=1\text{, }\varphi=\frac{\pi}{6}$, ReMAC $L=2\text{, }\varphi=\frac{\pi}{6}$, ReMAC $L=4\text{, }\varphi=\frac{\pi}{6}$, ReMAC $L=1\text{, }\varphi=\frac{\pi}{4}$, ReMAC $L=2\text{, }\varphi=\frac{\pi}{4}$, ReMAC $L=4\text{, }\varphi=\frac{\pi}{4}$, {\color{black}Digital }AirComp$\text{, }\varphi=\frac{\pi}{6}$};
\end{axis}
\end{tikzpicture}
  }\subfigure[$f=\max_{k}x_k,\sigma_z^2=1$]{\label{fig:fading_max_NMSE}
\centering
\begin{tikzpicture}[
    domain=0:4,
    spy using outlines={circle, magnification=3, size=2.3cm, connect spies, every spy on node/.append style={thick}}, 
    ]
    \begin{axis}[
        xlabel = {$\sigma_h^2$},
        ylabel = {NMSE},
        label style={font=\footnotesize},
        width=0.47\textwidth,
        height=7cm,
        xmin=0, xmax=0.5,
        ymin=0.0085, ymax=0.075,
        legend style={nodes={scale=0.65, transform shape}, at={(0.3,0.85)}},
        ticklabel style = {font=\footnotesize},
        legend pos=south east,
        ymajorgrids=true,
        xmajorgrids=true,
        grid style=dashed,
        grid=both,
        ymode = log,
        grid style={line width=.1pt, draw=gray!10},
        major grid style={line width=.2pt,draw=gray!30},
    ]
    \addplot[smooth,
             thin,
        color=airforceblue,
        mark=*,
        line width=0.5pt,
        mark size=2pt,
        ]
    table[x=sigmah,y=L1]
    {Data/Sim_fading_max.dat};
    \addplot[smooth,
              thin,
            color=chestnut,
            mark=square*,
            line width=0.5pt,
            mark size=2pt,
            ]
    table[x=sigmah,y=L2]
    {Data/Sim_fading_max.dat};
    \addplot[smooth,
             thin,
        color=cssgreen,
        mark=triangle*,
        line width=0.5pt,
        mark size=2pt,
        ]
    table[x=sigmah,y=L4]
    {Data/Sim_fading_max.dat};
    \addplot[smooth,
             thin,
             densely dashed,
        color=airforceblue,
        mark=*,
        line width=0.5pt,
        mark size=2pt,
        ]
    table[x=sigmah,y=L1p]
    {Data/Sim_fading_max.dat};
    \addplot[ smooth,
              thin,
            densely dashed,
            color=chestnut,
            mark=square*,
            line width=0.5pt,
            mark size=2pt,
            ]
    table[x=sigmah,y=L2p]
    {Data/Sim_fading_max.dat};
    \addplot[ smooth,
             thin,
        densely dashed,
        color=cssgreen,
        mark=triangle*,
        line width=0.5pt,
        mark size=2pt,
        ]
    table[x=sigmah,y=L4p]
    {Data/Sim_fading_max.dat};
    \addplot[ smooth,
             thin,
        densely dashed,
        color=cadmiumorange,
        mark=diamond*,
        line width=0.5pt,
        mark size=2pt,
        ]
    table[x=sigmah,y=aircomp]
    {Data/Sim_fading_max.dat};
\legend{ReMAC $L=1\text{, }\varphi=\frac{\pi}{6}$, ReMAC $L=2\text{, }\varphi=\frac{\pi}{6}$, ReMAC $L=4\text{, }\varphi=\frac{\pi}{6}$, ReMAC $L=1\text{, }\varphi=\frac{\pi}{4}$, ReMAC $L=2\text{, }\varphi=\frac{\pi}{4}$, ReMAC $L=4\text{, }\varphi=\frac{\pi}{4}$, {\color{black}Digital }AirComp$\text{, }\varphi=\frac{\pi}{6}$};
\end{axis}
\end{tikzpicture}
}
  \caption{Performance comparison between \ac{ReMAC}, ChannelComp and {\color{black}Digital} \ac{AirComp}  among $K=8$ nodes in the presence of fading channels under different transmission slots, i.e., $L=\{1,2,4\}$, and phase shifts, i.e., $\varphi=\{\pi/6,\pi/4\}$. Input values are $x_k \in \{1,2,3,4\}$. The \ac{NMSE} is depicted versus the variance of channel coefficients $\sigma_h^2$ for product function $f = \prod_{k=1}^Kx_k$ with low noise variance, i.e., {\color{black}$\sigma_z^2=0.1$}, and maximum function $f=\max_k x_k$ with high noise variance, i.e., $\sigma_z^2=1$, respectively. 
  }\label{fig:four_node_prod_NMS}
\end{figure*}

\begin{figure}[!t]
\centering
\begin{tikzpicture}[
    domain=0:4,
    spy using outlines={circle, magnification=3, size=1.5cm, connect spies, every spy on node/.append style={thick}}, 
    ]
    \begin{axis}[
        xlabel = {SNR},
        ylabel = {NMSE},
        label style={font=\footnotesize},
        width=0.47\textwidth,
        height=7cm,
        xmin=-11, xmax=13,
        ymin=0.045, ymax=0.3,
        legend style={nodes={scale=0.65, transform shape}, at={(0.3,0.85)}},
        ticklabel style = {font=\footnotesize},
        legend pos=south west,
        ymajorgrids=true,
        xmajorgrids=true,
        grid style=dashed,
        grid=both,
        ymode = log,
        grid style={line width=.1pt, draw=gray!10},
        major grid style={line width=.2pt,draw=gray!30},
    ]
    \addplot[smooth,
             thin,
        color=chestnut,
        mark=*,
        line width=0.5pt,
        mark size=2pt,
        ]
    table[x=SNR,y=L1]
    {Data/Sim_fading_bit_slicing.dat};
    \addplot[smooth,
              thin,
            color=airforceblue,
            mark=square*,
            line width=0.5pt,
            mark size=2pt,
            ]
    table[x=SNR,y=L2]
    {Data/Sim_fading_bit_slicing.dat};
    \addplot[ smooth,
             thin,
        color=cssgreen,
        mark=triangle*,
        line width=0.5pt,
        mark size=2pt,
        ]
    table[x=SNR,y=slicing2]
    {Data/Sim_fading_bit_slicing.dat};
\legend{ReMAC $L=1$, ReMAC $L=2$, Bit-Slicing $L=2$, Bit-Slicing $L=4$};
\end{axis}
\end{tikzpicture}
\caption{A performance comparison is conducted among ReMAC, ChannelComp, and Bit-Slicing across a network of \( K = 6 \) nodes in the presence of fading channels, where the fading variance is set to \( \sigma_h^2 = 1 \) and the phase shift is \( \varphi = \pi/4 \). The NMSE is evaluated against various SNR values and different numbers of transmission slots, i.e., \( L \in \{1,2\} \). The input values are \( x_k \in \{1,2,\ldots,64\} \), and the target function is the summation \( f = \sum_{k=1}^K x_k \).  }
\label{fig:six_node_sum_low_variance}
\end{figure}
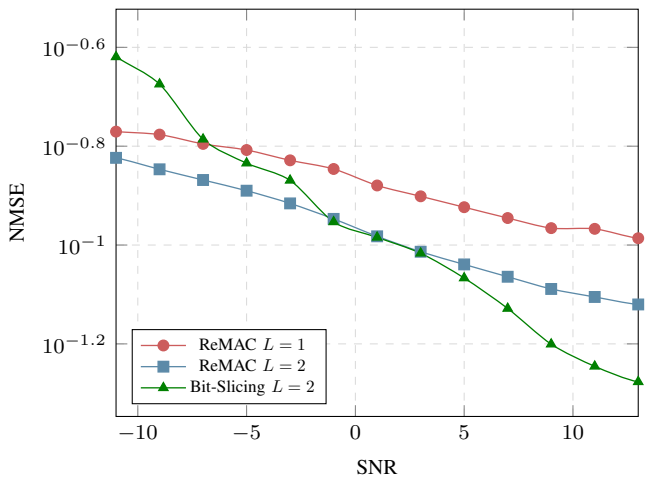

\subsection{Performance of \ac{ReMAC}} \label{subsec:performance of remac}

In this subsection, we first evaluate the performance of \ac{ReMAC} by solving Problem \(\mathcal{P}_0\) with Algorithm~\ref{Alg:HIT}. The analysis focuses on a network with \( K=4 \) nodes, where the input values \( x_k \in \{1, 2, \ldots, 256\} \) are represented by $8$ bits. The \ac{NMSE} performance of \ac{ReMAC} is assessed across different numbers of transmission time slots, \( L = \{1, 2, 3, 4\} \), and the results are averaged over \( N_s = 100 \) samples generated through Monte Carlo simulations.

Figure~\ref{fig:four_node_product_SN} shows the \ac{NMSE} of the function computation for the sum, product and maximum functions over different \ac{SNR} levels, respectively. As expected, increasing the \ac{SNR} leads to a decrease in \ac{NMSE} for all the functions. The most significant reduction occurs between $25$ dB and $40$ dB, regardless of the number of time slots.
{\color{black}Additionally, the \ac{NMSE} decreases with increasing time slots \(L\), as \ac{ReMAC} improves computational reliability by repeating modulated symbols, though at the cost of higher latency.} There are two notable cases for the assigned values of $L$. One case is when $L=4$, \ac{ReMAC} assigns each node a unique time slot for transmission, effectively operating as the orthogonal resource allocation technique known as \ac{TDMA}, and achieving the lowest \ac{NMSE}. The other is when $L=1$, \ac{ReMAC} simplifies to ChannelComp, resulting in a slight increase in \ac{NMSE}.

{\color{black}
Next, we proceed to evaluate the optimality gap through numerical experiments and the theoretical analysis in Proposition~\ref{theorem:convergence rate}. The approximate solution \( \bm{c}_{\ell}^{n} \) is obtained using Algorithm~\ref{Alg:HIT}, while the optimal solution \( \bm{c}_\ell^* \) is determined via exhaustive search for subproblem \( \mathcal{P}_2 \) at each iteration. Since the modulation vector \( \bm{x} \) lies in a continuous complex domain, the exhaustive search is applied only to \( \mathcal{P}_2 \).  
Given that \( \mathcal{P}_2 \) is a \ac{MIQCP} problem, the empirical optimality gap is measured as \( \sum_{\ell=1}^L\|\bm{c}_{\ell}^{n}\|_1 - \sum_{\ell=1}^L\|\bm{c}_{\ell}^{*}\|_1 \). Correspondingly, based on Proposition 2, we solely focus on the upper bound for the gap between the obtained and optimal solutions for \( \mathcal{P}_2 \), given by \( 2R_{2}(\bm{C},\bm{C}^*)/(n-1) \).  
The evaluation considers network sizes from \( K=3 \) to \( K=8 \) and quantization levels \( Q \in \{8, 16\} \), with the number of time slots fixed at \( L=2 \) for the product function. Algorithm~\ref{Alg:HIT} runs for \( n=20 \) iterations when \( Q=8 \) and \( n=30 \) iterations when \( Q=16 \). As shown in Figure~\ref{fig:optimality_gaps}, both the numerical and analytical gaps increase approximately linearly with network size, confirming that the proposed proximity analysis provides a valid upper bound for the observed optimality gap in the numerical experiments.}

\subsection{Comparison to {\color{black}Digital} AirComp, ChannelComp and {\color{black}Bit-slicing}}

{\color{black}In this subsection, we first evaluate the \ac{NMSE} performance of our proposed \ac{ReMAC} in comparison with digital \ac{AirComp}, ChannelComp, and bit-slicing for the summation function. Additionally, we compare \ac{ReMAC} with digital \ac{AirComp} and ChannelComp for the product and maximum.} 
The same power budget is allocated across all time slots for these methods to ensure a fair comparison.
Furthermore, we evaluate the effect of fading channels on the performance of \ac{ReMAC} in the absence of perfect \ac{CSI} at each node, where the channel coefficients $h_k$ are subject to random variations in both magnitude and phase. Specifically, the magnitude of the channel coefficient is generated with a normal Gaussian distribution, i.e., $h_k \sim \mathcal{N}(1, \sigma_h^2)$. In contrast, the phase of the channel coefficient is generated according to a uniform distribution, i.e., $\psi_k \sim \mathcal{U}(-\varphi, \varphi)$.

In Figure~\ref{fig:four_node_sum_NMS}, we compare the performance of \ac{ReMAC}, ChannelComp and digital \ac{AirComp} under various levels of noise and fading when computing the summation function. The input $x_k$ takes values from the set $\{1, 2, 3, 4\}$, and the evaluation is conducted over a network with $K = 8$ nodes. {\color{black} For digital \ac{AirComp}, the signals are averaged over \( L = 2 \) time slots.} Figure~\ref{fig:four_node_sum_NMSE} shows the \ac{NMSE} performance over different fading variances for various numbers of time slots $L = \{1, 2, 4\}$ and phase shifts {\color{black}$\varphi = \{\pi/4, \pi/2\}$} under a low noise scenario, i.e., 
$\sigma_z^2=0.1$.
It can be observed that a larger number of time slots results in a lower \ac{NMSE}, which is consistent with the results in Figure \ref{fig:four_node_product_SN}. This finding highlights that \ac{ReMAC} also exhibits superior performance under fading channels compared to ChannelComp, which has already been shown to outperform digital \ac{AirComp}~\cite{razavikia2023computing}. 
As expected, when the fading variance or phase shift increases, the transmitted symbols experience greater distortion from the original ones, resulting in a higher \ac{NMSE}. However, once the fading variance exceeds 0.6, the rate of increase in \ac{NMSE} slows down, and further increases in phase shift have minimal impact on \ac{NMSE}.

Figure~\ref{fig:four_node_prod_NMSE} shows a similar curve in a high noise scenario with {\color{black}$\sigma_z^2=1$}. Due to the increased channel noise variance, the \ac{NMSE} becomes higher than that in Figure~\ref{fig:four_node_sum_NMSE}. Nevertheless, \ac{ReMAC} consistently outperforms ChannelComp and digital AirComp across various fading conditions.

In the next experiment, shown in Figure~\ref{fig:four_node_prod_NMS}, we compare the performance of \ac{ReMAC}, ChannelComp, and digital \ac{AirComp} for computing the product and maximum functions among \( K=8 \) nodes under different channel fading variances. The input values \( x_k \) are selected from the set \( \{1, 2, 3, 4\} \), and the number of transmission time slots $L$ is varied across \( \{1, 2, 4\} \). Figure~\ref{fig:fading_prod_NMSE} shows that in a low noise scenario, i.e., {\color{black}$\sigma_z^2=0.1$}, \ac{ReMAC} outperforms ChannelComp across various fading levels when computing the product function. Specifically, \ac{ReMAC} reduces the computation error by approximately $7.5$~dB compared to ChannelComp under low fading variance. As fading variance increases, the performance of \ac{ReMAC} saturates, and the \ac{NMSE} difference between varying numbers of time slots narrows down. Moreover, since {\color{black}digital} \ac{AirComp} approximates the product using the log function~\cite{csahin2023survey}, it cannot compute accurately even in the low noise scenario.

Similarly, Figure~\ref{fig:fading_max_NMSE} shows that in a high noise scenario, i.e., $\sigma_z^2=1$, \ac{ReMAC} outperforms ChannelComp in computing the maximum function, reducing the \ac{NMSE} by approximately $1.5$ dB. However, increasing the time slots in this high noise scenario does not significantly reduce the \ac{NMSE}. Additionally, {\color{black}digital} \ac{AirComp} uses an exponential function to approximate the maximum~\cite{csahin2023survey}, hence resulting in a higher \ac{NMSE}.
Overall, these numerical results show that by jointly designing the constellation diagram and the {\color{black}coded repetition}, \ac{ReMAC} outperforms ChannelComp and {\color{black}digital} \ac{AirComp} over both noisy and fading channels, particularly for non-summation functions. 

{\color{black} 
Finally, in Figure~\ref{fig:six_node_sum_low_variance}, we analyze the performance of the summation function by comparing \ac{ReMAC}, ChannelComp, and bit-slicing for input values \( x_k \in \{1, 2, \ldots, 64\} \) in a network with \( K=6 \) nodes. The fading channel variance is set to \( \sigma_h^2=1 \), and the phase shift is \( \varphi=\pi/4 \). The transmission time slots are chosen from \( L \in \{1, 2\} \). Across a range of SNR values from \([-10, 10]\), \ac{ReMAC} demonstrates a lower NMSE compared to ChannelComp, aligning with the trends observed in Figure~\ref{fig:four_node_product_SN}. 
The results further indicate that bit-slicing outperforms \ac{ReMAC} in high-SNR conditions. However, as the SNR decreases, bit-slicing surpasses \ac{ReMAC}, which becomes more vulnerable to disruptions. Specifically, for bit-slicing, errors in decoding a sliced integer over consecutive \( L \) time slots due to fading distortions and noise directly lead to computation errors in the function output. In contrast, \ac{ReMAC} employs a joint modulation and coded repetition approach, where the \ac{CP} compares the aggregated symbol vector across \( L \) time slots against modulation vectors in the codebook. Even if a symbol in a particular time slot experiences severe distortion, the redundancy introduced by repetition helps recover the function output more accurately. This design enables \ac{ReMAC} to effectively mitigate noise and fading distortions, ensuring more reliable computation under low-SNR conditions. 
Moreover, increasing the number of transmission slots \( L \) improves \ac{NMSE} performance for both \ac{ReMAC} and bit-slicing. This enhancement also highlights the tradeoff between achieving higher computational accuracy and balancing transmission latency.
}

\section{Conclusion}\label{sec:conclusion}

This paper introduced \ac{ReMAC}, a joint modulation and coding scheme for repeated transmission in digital over-the-air computation, to provide reliable communication for computation. Building upon the ChannelComp framework, we designed the {\color{black}coded repetition} scheme to reduce the computation error over \ac{MAC}. To this end, we proposed an optimization problem that jointly determines the encoding for digital modulation over multiple time slots. To manage the computational complexity of the proposed optimization problem, we developed an alternating minimization approach. Moreover, we evaluated the effectiveness of \ac{ReMAC} through the numerical experiment by comparing it to existing state-of-the-art methods, such as \ac{AirComp} and ChannelComp. Notably, we observed approximately $7.5$~dB improvement in reducing the \ac{NMSE} of the computation error for the product function in the presence of fading.

Further exploration of \ac{ReMAC} could proceed in various directions, including but not limited to the following:
\begin{itemize}
    \item \textbf{Other optimization perspectives}: Machine learning techniques could be explored to implement the modulation and channel code design by treating it as a learning task, making it possible to achieve accurate function computation more efficiently than conventional convex optimization-based methods.
    \item \textbf{Integrated with distributed learning}: We plan to integrate \ac{ReMAC} with distributed learning to enable efficient and scalable aggregation of model updates in a distributed setting. By introducing the co-design of modulation and repeated transmission, we aim to enhance the efficiency and accuracy of distributed learning tasks. 
    \item \textbf{MIMO extension for parallel computation}: We intend to upgrade the current single narrowband antenna network to a broadband multiple-input and multiple-output network, enabling parallel computations to improve the computation throughput for several applications, such as distributed estimation or distributed data analytics.
    \item {\color{black}\textbf{Optimization for minimizing NMSE}: One interesting direction to explore is considering NMSE minimization by leveraging modulation schemes such as \ac{PAM} and \ac{QAM} and designing repetition patterns across consecutive time slots under certain transmission budget constraints.}
    \item {\color{black}\textbf{Phase-aligned transmission with precoding errors}: We aim to extend \ac{ReMAC} to address the challenges of achieving phase-aligned precoded transmission in practical scenarios. Future work will explore techniques such as pre-equalization, and incorporate more realistic models with precoding errors in real-world deployments.}
\end{itemize}

\appendix

\subsection{Proof of Proposition~\ref{pro:linear relaxation}} \label{app:linear relaxation}

It is evident that Problem $\mathcal{P}_2$ can be reformulated as the following equivalent problem:
\begin{align}
\nonumber
 \bm{c}_{\ell} = \arg & \min_{\bm{c}_{\ell}} \sum_{\ell=1}^L \|\bm{c}_{\ell}\|_1, \\
\text{s.t.} & \sum_{\ell=1}^L \bm{c}_{\ell}^{\top} \bm{P}_{i,j} \bm{c}_{\ell} \leq -\Delta f_{i,j}, \label{eq:P2_replace}
\end{align}
where $\bm{P}_{i,j} :=(\bm{a}_i-\bm{a}_j) (\bm{a}_j-\bm{a}_i)^{\top} \odot \bm{W}^n$, and $\bm{W}^n = \bm{x}^n\bm{x}^{n \mathsf{H}}$. Hence, the matrix  $\bm{P}_{i,j}$ is a symmetric matrix with ${\rm rank}(\bm{P}_{i,j})=r_{ij}$, and we can decompose it as $\bm{P}_{i,j} = \sum_{m=1}^{r_{ij}}\bm{p}_{m}^{i,j} \bm{p}_{m}^{i,j \top}$ for any vector $\bm{p}_{m}^{i,j} \in \mathbb{R}^{N}$. 
By substituting the decomposition into  \eqref{eq:P2_replace}, we obtain the following optimization 
\begin{align}
\nonumber
 \bm{c}_{\ell} = \arg & \min_{\bm{c}_{\ell}} \sum_{\ell=1}^L \mathds{1}^{\top}\bm{c}_{\ell}, \\
\text{s.t.} & \sum_{\ell=1}^L\sum_{m=1}^{r_{ij}}\left(\bm{p}_{m}^{i,j\top }\bm{c}_{\ell}\right)\left(\bm{p}_{m}^{i,j\top }\bm{c}_{\ell}\right) \leq -\Delta f_{i,j}. \label{eq:MBO_reformulated3}
\end{align}
Here, we also used the fact that entries of vector $\bm{c}_{\ell}$ are positive values for the cost function.  Then, we relax the binary restriction on $\bm{c}_{\ell}$ for Optimization in \eqref{eq:MBO_reformulated3}, resulting in a continuous feasible region $\mathcal{C}$. The continuous feasible region $\mathcal{C}$ is defined as
\begin{align*}
\mathcal{C} := \{\bm{c}_\ell \in [0,1]^N, \forall \ell \in [L]\},
\end{align*}
where $L$ represents the number of time slots. Thereafter, let $l_{m}^{i,j}$ and  $u_{m}^{i,j}$ be the values as
\begin{subequations} \label{eq:bilinear_bound2}
\begin{align}
l_{m}^{i,j} &= \sum_{n=1}^N\nolimits\min\{0,p_{m,n}^{i,j}\}, \\
u_{m}^{i,j} &= \sum_{n=1}^N\nolimits\max\{0,p_{m,n}^{i,j}\}, 
\end{align}
\end{subequations}
where $\forall m \in [r_{ij}]$ and $p_{m,n}^{i,j}$ is the $n$-th element of $\bm{p}_{m}^{i,j}$.  To deal with the non-linear optimization with bilinear constraints, we use a well-known piecewise McCormick-based relaxation technique~\cite{miro2012deterministic}. Based on the defined minimum and maximum values from \eqref{eq:bilinear_bound2}, we can derive the following inequalities.
\begin{subequations}
\label{eq:bilinear_bound_inequality}
\begin{align}
(\bm{p}_{m}^{i,j \top}\bm{c}_{\ell}-l_{m}^{i,j})^2  & \geq 0, \\ 
(\bm{p}_{m}^{i,j \top}\bm{c}_{\ell}-u_{m}^{i,j})^2  & \geq 0.
\end{align}
\end{subequations}

To this end, a lower bound for each bilinear term in the constraint in \eqref{eq:MBO_reformulated3},  following the conclusion from \eqref{eq:bilinear_bound_inequality}, can be obtained as
\begin{align} \nonumber
h_{\ell}^{i,j}(\bm{C}) 
=&\sum_{m=1}^{r_{ij}} (\bm{p}_{m}^{i,j\top }\bm{c}_{\ell})^2, \\ \nonumber
\geq & \max \Big\{ \sum_{m=1}^{r_{ij}} 2l_{m}^{i,j} \bm{p}_{m}^{i,j \top}\bm{c}_{\ell}-(l_{m}^{i,j})^2, \\  \nonumber
& \sum_{m=1}^{r_{ij}}  2u_{m}^{i,j} \bm{p}_{m}^{i,j \top}\bm{c}_{\ell}-(u_{m}^{i,j})^2\Big\}
\triangleq  \hat{h}_{\ell}^{i,j}(\bm{C}).
\end{align}

Consequently, the \ac{MIQCP} in optimization problem in \eqref{eq:P2_replace} can be rewritten as a linear relaxation programming problem as follows:
\begin{subequations}
\label{eq:relaxation_programming}
\begin{align}
&\mathcal{P}_{4} :=  \min_{\bm{c}_{\ell}}  \quad \sum_{\ell=1}^L \mathds{1}^{\top}\bm{c}_{\ell}, \nonumber \\
\text{s.t.} & \sum_{\ell=1}^L \sum_{m=1}^{r_{ij}} 
2l_{m}^{i,j}\bm{p}_{m}^{i,j\top}\bm{c}_{\ell} \leq -\Delta f_{i,j} + \sum_{\ell=1}^L \sum_{m=1}^{r_{ij}}(l_{m}^{i,j})^2, \\
& \sum_{\ell=1}^L\sum_{m=1}^{r_{ij}} 
2u_{m}^{i,j}\bm{p}_{m}^{i,j\top}\bm{c}_{\ell} \leq -\Delta f_{i,j} + \sum_{\ell=1}^L \sum_{m=1}^{r_{ij}}(u_{m}^{i,j})^2, \\
& \bm{c}_\ell \in [0,1]^N, \quad \forall \ell \in [L].
\end{align}
\end{subequations}

Furthermore, for the optimality gap, given any $\bm{c}_{\ell}$, we can find out that 
\begin{align} \nonumber
\Delta h_{\ell}^{i,j}(\bm{C})&: =h_{\ell}^{i,j}(\bm{C}) - \hat{h}_{\ell}^{i,j}(\bm{C}),\\ \nonumber
= &\max \{\sum_{m=1}^{r_{ij}}\bm{p}_{m}^{i,j\top} \bm{c}_{\ell} \bm{p}_{m}^{i,j\top} \bm{c}_{\ell} - 2l_m^{i,j}\bm{p}_{m}^{i,j\top}\bm{c}_{\ell}+(l_{m}^{i,j})^2,\\ \nonumber
&\qquad \sum_{m=1}^{r_{ij}}\bm{p}_{m}^{i,j\top} \bm{c}_{\ell} \bm{p}_{m}^{i,j\top} \bm{c}_{\ell} - 2u_{m}^{i,j}\bm{p}_{m}^{i,j\top}\bm{c}_{\ell}+(u_{m}^{i,j})^2\}, \\ \nonumber
= & \max \{\sum_{m=1}^{r_{ij}}\left(\bm{p}_{m}^{i,j\top} \bm{c}_{\ell}-l_{m}^{i,j}\right)\left(\bm{p}_{m}^{i,j\top} \bm{c}_{\ell}-l_{m}^{i,j}\right), \\ \nonumber
&\qquad \quad \sum_{m=1}^{r_{ij}}\left(\bm{p}_{m}^{i,j\top} \bm{c}_{\ell}-u_{m}^{i,j}\right)\left(\bm{p}_{m}^{i,j\top} \bm{c}_{\ell}-u_{m}^{i,j}\right)\},\\ \nonumber
\leq & \max \{\sum_{m=1}^{r_{ij}}\left(u_{m}^{i,j} - l_{m}^{i,j} \right)^2, 
\sum_{m=1}^{r_{ij}}\left(u_{m}^{i,j} - l_{m}^{i,j} \right)^2\}, \\
= & \|\bm{u}^{i,j}-\bm{l}^{i,j}\|_2^2,
\label{eq:bilinear_objective_inequality_gap}
\end{align}
where $\bm{l}^{i,j}:=[l_1^{i,j},\ldots,l_{r_{ij}}^{i,j}]$ and $\bm{u}^{i,j}:=[u_1^{i,j},\ldots,u_{r_{ij}}^{i,j}]$. Hence, by narrowing down the value of each entry of $\bm{c}_{\ell}$, we can obtain $\Delta h_{\ell}^{i,j}(\bm{C}) \rightarrow 0$, accordingly, the sum values $\sum_{\ell=1}^L\Delta h_{\ell}^{i,j}(\bm{C}) \rightarrow 0$. As a result, the constraints in Problem $\mathcal{P}_4$ approaches constraints in Problem $\mathcal{P}_2$, which concludes the proof.

\subsection{Branch and Bound for Solving $\mathcal{P}_2$}\label{app:MIQCPconvergence}

To obtain the optimal solution for Problem $\mathcal{P}_2$ by solving its relaxed version Problem $\mathcal{P}_4$, the branch and bound method can be employed to iteratively reduce the optimality gap between these two problems.
More precisely, let $\mathcal{C}^{t}$ be the union of all the feasible regions for Problem $\mathcal{P}_4$ at iteration $t$, where each entry of $\bm{c}_{\ell}$ belongs to the interval between zero and one, i.e., $c_{\ell,n}\in [0,1]$ for  $\ell \in [L]$ and  $n\in [N]$.  After solving Problem $\mathcal{P}_4$, we obtain the {\color{black} repetition coding matrix} $\bm{C}^{(t)*} =[\bm{c}_1^{(t) *},\ldots, \bm{c}_K^{(t) *}]$ as optimal solution to the relaxed $\mathcal{P}_4$. Note that by substituting $\bm{C}^{(t)*}$ into the objective function of $\mathcal{P}_4$, we obtain a lower bound for the optimal objective value of $\mathcal{P}_2$, and we denote this lower bound as ${\rm LB}^{(t)}$. Also, we can obtain an upper bound for the optimal objective function value of $\mathcal{P}_2$ by substituting $\bm{C}^{(t)*}$  into $\mathcal{P}_2$ denoted by ${\rm UB}^{(t)}$. 
Now, we use these lower and upper bounds in every iteration to reach the optimal solution for Problem $\mathcal{P}_2$. In this regard, let $\delta >0$ be the convergence tolerance and ${\rm UB}^{(t)} - {\rm LB}^{(t)} \leq \delta$ be the termination criteria.

At iteration $t$, if all the entries of $\bm{C}^{(t)*}$ are binary, then  $\bm{C}^{(t)*}$ is the optimal solution to $\mathcal{P}_2$ and the procedure can stop. Otherwise,  for the entries of $\bm{C}^{(t)*}$ that are not binary, we need to search by resolving the  $\mathcal{P}_4$ for both cases of zero and one. For instance, consider that $c_{\ell, n'}^{(t)*}$ is one non-binary entry of the solution to $\mathcal{P}_4$ at iteration $t$ for some $\ell$ and $n'$, we partition $\mathcal{C}^{t}$ into (at least) two new sub-regions such that $\mathcal{C}^{t}_0$ and  $\mathcal{C}^{t}_1$ corresponds to $c_{\ell, n'}^{(t)*} = 0$ and $c_{\ell, n'}^{(t)*} = 1$, respectively.  Afterward, we solve $\mathcal{P}_4$ over the feasible set $\mathcal{C}^{t}_0$ and $\mathcal{C}^{t}_1$  with the corresponding solutions $\hat{c}_{\ell, n'}^{0}$ and $\hat{c}_{\ell, n'}^{1}$, respectively. Then, we substitute these solutions into Problems $\mathcal{P}_2$ and $\mathcal{P}_4$ to obtain the new lower and upper bounds, respectively. Indeed, considering the optimum values $s_0^t:= \mathcal{P}_2(\hat{c}_{\ell, n'}^{0})$,  $s_1^t:= \mathcal{P}_2(\hat{c}_{\ell, n'}^{1})$ and $r_0^t:= \mathcal{P}_4(\hat{c}_{\ell, n'}^{0})$,  $r_1^t:= \mathcal{P}_4(\hat{c}_{\ell, n'}^{1})$, we can update the lower and upper bounds as follows.
\begin{align}
     {\rm UB}^{(t+1)} & = \min\{{\rm UB}^{(t)}, r_0^t, r_1^t\}, \\
     {\rm LB}^{(t+1)} & = \max\{{\rm LB}^{(t)}, s_0^t, s_1^t\}.
\end{align}

We continue this procedure until either the solution $\bm{c}_{\ell}^{(t)*}$ becomes completely a binary vector or the gap between the global lower and upper bound becomes small enough, i.e.,  ${\rm UB}^{(t)} - {\rm LB}^{(t)} \leq \delta$. 

By following the proposed procedure, in~\cite{zhao2017global}, they showed that we can obtain an optimal solution to $\mathcal{P}_2$ within finite iterations. In particular, the following theorem guarantees the convergence to the optimal solution.

\begin{theorem}[\!\!{\cite[Theorem 2]{zhao2017global}}]\label{the:Branch} The proposed algorithm~terminates within finite iterations by an optimal solution for $\mathcal{P}_2$ or by an evidence indicating that Problem $\mathcal{P}_2$ is infeasible.
\end{theorem}

Theorem~\ref{the:Branch} implies that the solution obtained from the branch and bound method provides a tight lower bound for the optimal solution of Problem $\mathcal{P}_2$.

\subsection{Proof of Proposition~\ref{theorem:convergence rate}} \label{app:convergence rate}

The alternating between optimization Problems $\mathcal{P}_3$ and $\mathcal{P}_4$ can be seen as a standard alternating minimization approach in which, at each iteration, we minimize an implicit objective function with respect to variables $\bm{c}_{\ell}$s and $\bm{W}$. To be more precise, let us define an indicator function ${\rm id}_{\mathcal{S}}(\bm{x})$  for $\mathcal{S}$, which is a closed subset of $\mathbb{R}^{n}$, as follows:
\begin{align}
 {\rm id}_{\mathcal{S}}(\bm{x}) = \begin{cases}
     0 \quad \text{if}~\bm{x} \in \mathcal{S}, \\
     +\infty \quad \text{otherwise}.
 \end{cases}
\end{align}

Furthermore, let $\mathcal{S}_3^n$ and $\mathcal{S}_4^n$ denote the feasible regions of Problem $\mathcal{P}_3$ and Problem $\mathcal{P}_4$ at iteration $n$, respectively.
\begin{align*}
    \mathcal{S}_3^{n}& := \{\bm{W}~| {\rm Tr}(\bm{W} \cdot \bm{B}_{i,j}^{n-1}) \geq \Delta f_{i,j},  \bm{W} \succeq \bm{0},~\text{Tr}(\mathbf{W}) \leq 1 \}, \\
    \mathcal{S}_4^{n}& := \{\bm{c}~| \mathcal{A}_{i,j}(\bm{c}, \bm{W}^{n})\leq 0 , \bm{c} \in [0,1]^{N}\},
\end{align*}
where operator $ \mathcal{A}_{i,j}$s represent the linear constraints in \eqref{eq:relaxation_programming}.  
Then, consider the following objective function. 
\begin{align}
    \mathcal{F}(\bm{C},\bm{W}) := \sum_{\ell=1}^L \|\bm{c}_{\ell}\|_1 + \sum_{\ell=1}^{L}{\rm id}_{\mathcal{S}_4^n}(\bm{c}_{\ell}) + {\rm id}_{\mathcal{S}_3^n}(\bm{W}).
\end{align}

Therefore, the iterative procedure in Algorithm~\ref{Alg:HIT} can be expressed as follows.
\begin{subequations}
\label{eq:ALt_modeled}
\begin{align}
\bm{W}^{n} & = \arg \min_{\bm{W}}  \mathcal{F}(\bm{C}^{n-1},\bm{W}), \\
\bm{C}^{n} & = \arg \min_{\bm{C}} \mathcal{F}(\bm{C},\bm{W}^{n}).
\end{align}
\end{subequations}

Since both $\mathcal{S}_3^n$ and $\mathcal{S}_4^n$ are convex sets and $\sum_{\ell=1}^L \|\bm{c}_{\ell}\|_1 $ is the combination of convex functions,  then the objective function $\mathcal{F}(\bm{C},\bm{W})$ is bi-convex. Hence, the proposed method can be considered as a special case of successive global minimization methods, whose convergence rate is stated below.
\begin{theorem}[\!\!\!{\cite[Theorem 5.2]{beck2013convergence}}]
Let $(\bm{W}^{n}, \bm{C}^{n})$ be the sequence generated by the alternating minimization approach in \eqref{eq:ALt_modeled}. Then, 
\begin{align}
\label{eq:Conv_rate}
\mathcal{F}(\bm{C}^{n},\bm{W}^{n}) - \mathcal{F}(\hat{\bm{C}},\hat{\bm{W}}) \leq \frac{2 \min\{L_1,L_2\} \tilde{R}^2(\bm{C}^0,\bm{W}^{0})}{n-1},
 \end{align}
for $n\geq 2$, $L_1$ and $L_2$ are the Lipschitz constants of $\mathcal{F}(\bm{C},\bm{W})$ respect to $\bm{W}$ and $\bm{C}$. Also $\tilde{R}(\bm{C}^0,\bm{W}^{0})$ is defined below:
    \begin{align*}   \tilde{R}(\bm{C}^0,\bm{W}^{0}):=\max_{\substack{(\bm{W},\bm{C})\\\in \mathcal{S}}} \max_{\substack{(\hat{\bm{W}},\hat{\bm{C}})\\\in \hat{\mathcal{S}}}} \Big\{&\sqrt{\|\bm{W}-\hat{\bm{W}}\|_{\rm F}^2+\|\bm{C}-\hat{\bm{C}}\|_{\rm F}^2}, \\
    & \mathcal{F}(\bm{C},\bm{W}) \leq \mathcal{F}(\bm{C}^0,\bm{W}^{0}) \Big\}. 
    \end{align*}
\end{theorem}
To obtain the Lipschitz constants, we need to obtain the gradient $\mathcal{F}(\bm{C},\bm{W})$ for the variables. When optimizing with respect to $\bm{W}$, $\mathcal{F}(\bm{C},\bm{W})$ becomes an indicator function, resulting in a Lipschitz constant of one, i.e., $L_1 = 1$. Additionally, since the $\ell_1$ norm is $1$-Lipschitz, the sum $\sum_{\ell=1}^L  \|\bm{c}_{\ell}\|_1$ becomes $L$-Lipschitz, hence yielding $L_2 = L$.

After alternating between $\mathcal{P}_3$ and $\mathcal{P}_4$, we use the branch and bound method to obtain the binary variables $\hat{\bm{c}}_{\ell}$ as the global minimizer for $\mathcal{P}_2$. To this end, the obtained solution $(\hat{\bm{W}},\hat{\bm{C}})$ satisfies
the constraints in both $\mathcal{P}_3$ and $\mathcal{P}_4$, meaning that  ${\rm id}_{\mathcal{S}_3}(\hat{\bm{W}})$ and ${\rm id}_{\mathcal{S}_4}(\hat{\bm{c}}_{\ell})$ become zero.
As a result,  $\mathcal{F}(\hat{\bm{C}},\hat{\bm{W}})$ becomes $\sum_{\ell=1}^L\|\hat{\bm{c}}_{\ell}\|_1$.  Next, by substituting $L_1 =1$ and $L_2 = L$ into  \eqref{eq:Conv_rate}, we have 
 \begin{align}
     \label{eq:cl_clhat}
     \sum_{\ell=1}^L  \|\bm{c}_{\ell}^{n}\|_1 -  \sum_{\ell=1}^L  \|\hat{\bm{c}}_{\ell}\|_1 \leq \frac{2 \tilde{R}^2(\bm{C}^0,\bm{W}^{0})}{n-1},
 \end{align}
where we use the fact that the obtained solution $\bm{W}^n$ and $\bm{c}_{\ell}^n$ satisfy the constrains in \eqref{eq:LiftSDPconvex} and \eqref{eq:relaxation_programming_high}, respectively, i.e., $\mathcal{F}(\bm{C}^n,\bm{W}^n)=\|\bm{c}_{\ell}^{n}\|_1$.
From~\cite{liu2023fast}, we know that $\|\hat{\bm{c}}_{\ell}\|_1 \leq \|\bm{c}_{\ell}^{*}\|_1 $, where $\bm{c}_{\ell}^{*}$ denotes the optimal point of $\mathcal{P}_2$. Moreover, the solution to $\mathcal{P}_3$ provides a lower bound for the optimal solution to $\mathcal{P}_1$ as expressed below~\cite{luo2010sdp}.
\begin{align}
    \| \hat{\bm{W}} \|_{\rm F}^2  \leq 8\tilde{M} \|{\bm{W}}^{*} \|_{\rm F}^2,
\end{align}
where $\tilde{M}$ indicates the number of constraints in \eqref{eq:CVX} and $\bm{W}^{*}$ represents the optimal point of $\mathcal{P}_1$. Let $\tilde{\bm{C}}$ and $\tilde{\bm{W}}$ be the feasible points that maximize $\tilde{R}(\bm{C}^{0}, \bm{W}^{0})$, i.e., the most distant feasible points from the obtained solution $\hat{\bm{C}}$ and $\hat{\bm{W}}$. Using the triangle inequality, we can derive an upper bound for the distance between $\tilde{\bm{W}}$ and $\hat{\bm{W}}$ as follows.
\begin{align}
   \nonumber
    \| \tilde{\bm{W}} - \hat{\bm{W}} \|_{\rm F}  &  \leq  \| \tilde{\bm{W}} - \bm{W}^* \|_{\rm F}  + \| \bm{W}^* - \hat{\bm{W}}\|_{\rm F},  \\ \nonumber
    & \leq \| \tilde{\bm{W}} - \bm{W}^* \|_{\rm F} + (1+2\sqrt{2\tilde{M}})\|\bm{W}^{*}\|_{\rm F}, \\
    & \leq \| \tilde{\bm{W}} - \bm{W}^* \|_{\rm F} + 3\sqrt{\tilde{M}}.
\end{align} 

The last inequality follows the fact that $\|\bm{W}^{*}\|_{\rm F}\leq 1$, as derived from \eqref{eq:LiftSDP}. Similarly, we can further upper bound the distance between $\tilde{\bm{W}}$ and $\bm{W}^*$, where $\bm{W}^*:=\bm{x}^{*}\bm{x}^{* \mathsf{H}}$.
\begin{subequations}\label{eq:Mtild}
\begin{align} 
   \nonumber
    \| \tilde{\bm{W}} - \hat{\bm{W}} \|_{\rm F}  \leq & \| \tilde{\bm{x}}\tilde{\bm{x}}^{\mathsf{H}} - \bm{x}^{*}\bm{x}^{* \mathsf{H}}\|_{\rm F} + \| \tilde{\bm{x}}\tilde{\bm{x}}^{\mathsf{H}}  - \tilde{\bm{W}} \|_{\rm F}  \\ \label{eq:Mtild_triangle} 
    & + 3\sqrt{\tilde{M}}, \\ \label{eq:Mtild_eigenvalue} 
    \leq & 2\|\tilde{\bm{x}} - \bm{x}^{*} \|_2^2 + \underbrace{\sum_{i=2}^N\tilde{\lambda}_i}_{\leq 1} + 3\sqrt{\tilde{M}}, 
    \\ \label{eq:Mtild_xdif}
    \leq & 2\|\tilde{\bm{x}}- \bm{x}^{*} \|_2^2 +  4\sqrt{\tilde{M}}.
\end{align}
\end{subequations}
Here, $\tilde{\lambda}_i, i \in [N]$ are the eigenvalues of the matrix $\tilde{\bm{W}}$, and $\tilde{\bm{x}}$ is the modulation vector obtained from $\tilde{\bm{W}}$ using the method described in \eqref{eq:Cholesky decomposition}. The inequality in \eqref{eq:Mtild_eigenvalue} is derived based on the fact that both $\tilde{\bm{x}}$ and $\bm{x}^{*}$ are unit vectors. By substituting \eqref{eq:Mtild} into \eqref{eq:cl_clhat}, we obtain the following upper bound.
 \begin{align} \nonumber
     \sum_{\ell=1}^L  \|\bm{c}_{\ell}^{n}\|_1 -  \sum_{\ell=1}^L  \|{\hat{\bm{c}}_{\ell}}\|_1 \leq & \frac{2}{n-1}(2\| \tilde{\bm{x}} - \bm{x}^{*} \|_2^2 + 4\sqrt{\tilde{M}})^2  \\
     & + \frac{2}{n-1}\sum_{\ell=1}^L\| \tilde{\bm{c}}_{\ell} -\hat{\bm{c}}_{\ell} \|_2^2.
 \end{align}

Additionally, by invoking the fact that $\bm{c}_{\ell}^{*}$ and $\hat{\bm{c}}_{\ell}$ are both binary vectors, we can get
\begin{align}
    \|\hat{\bm{c}}_{\ell}\|_2  = \|\hat{\bm{c}}_{\ell}\|_1\leq \|\bm{c}_{\ell}^{*}\|_1 = \|\bm{c}_{\ell}^{*}\|_2, ~\forall \ell \in [L].
    \label{eq:binary inequality}
\end{align}

Finally, by applying the triangle inequality to $\| \tilde{\bm{c}}_{\ell} -\hat{\bm{c}}_{\ell} \|_2$ and using the fact $\|\bm{c}_{\ell}^{*} 
 - \hat{\bm{c}}_{\ell}\|_2 \leq 2 \|\bm{c}_{\ell}^{*}\|_2$ derived from~\eqref{eq:binary inequality}, the convergence rate can be obtained as follows.
 \begin{align*}
     \sum_{\ell=1}^L  \|\bm{c}_{\ell}^{n}\|_1 -  \sum_{\ell=1}^L  \|{\bm{c}}_{\ell}^{*}\|_1 & \leq \frac{2}{n-1}\Big(2\| \tilde{\bm{x}} - \bm{x}^{*} \|_2^2 + 4\sqrt{\tilde{M}}\Big)^2  \\
     & + \frac{2}{n-1}\sum_{\ell=1}^L\Big(\|\tilde{\bm{c}}_{\ell} -\bm{c}_{\ell}^{*} \|_2 + 2 \|\bm{c}_{\ell}^{*}\|_2 \Big)^2,
 \end{align*}
which concludes the proof.

\bibliographystyle{IEEEtran}
\bibliography{Ref}

\end{document}